\documentclass[reprint,amsmath,amssymb,prl,superscriptaddress]{revtex4-2}

\usepackage{graphicx}
\usepackage{bm}
\usepackage{amsmath,amssymb}
\usepackage{xspace}
\usepackage{braket}
\usepackage{comment}
\usepackage{tabularx}
\usepackage{hyperref}
\hypersetup{
    colorlinks = true,
    allcolors = {blue}
}

\usepackage[dvipsnames]{xcolor}
\definecolor{dcyan}{RGB}{0,100,100}
\AtBeginDocument{\let\oldcontentsline\contentsline}
\newcommand{\notoccontentsline}[4]{\oldcontentsline{}{}{}{}}
\newcommand{\droptocpage}{\addtocontents{toc}{\let\protect\contentsline\protect\notoccontentsline}}
\newcommand{\incltocpage}{\addtocontents{toc}{\let\protect\contentsline\protect\oldcontentsline}}

\newcommand{\multiL}{\biggl[\biggr.}

\begin{document}
	\title{Quantum-limited millimeter wave to optical transduction} 
    
    \author{Aishwarya Kumar}
    \altaffiliation{The authors contributed equally to this work}
    \author{Aziza Suleymanzade}
    \altaffiliation{The authors contributed equally to this work}
    \author{Mark Stone}
    \altaffiliation{The authors contributed equally to this work}
    \author{Lavanya Taneja}
    \author{Alexander Anferov}
    \affiliation{The Department of Physics, The James Franck Institute, and The Pritzker School of Molecular Engineering, The University of Chicago, Chicago, IL}
    \author{David I. Schuster}
    \affiliation{The Department of Physics, The James Franck Institute, and The Pritzker School of Molecular Engineering, The University of Chicago, Chicago, IL}
    \affiliation{The Department of Applied Physics, Stanford University, Stanford, CA}
    \author{Jonathan Simon}
    \affiliation{The Department of Physics, The James Franck Institute, and The Pritzker School of Molecular Engineering, The University of Chicago, Chicago, IL}
    \affiliation{The Department of Applied Physics, Stanford University, Stanford, CA}
    \affiliation{The Department of Physics, Stanford University, Stanford, CA}


\date{\today}
	\begin{abstract} 

Long distance transmission of quantum information is a central ingredient of distributed quantum information processors for both computing and secure communication~\cite{kimble2008quantum}. Transmission between superconducting/solid-state quantum processors~\cite{arute2019quantum} necessitates transduction of individual microwave photons – the natural excitations of the processors, to optical photons – the natural low noise information carriers at room temperature~\cite{lauk2020perspectives}. Current approaches to transduction employ solid state links between electrical and optical domains, facing challenges from the thermal noise added by the strong classical pumps required for high conversion efficiency and bandwidth~\cite{mckenna2020cryogenic,Xu2021,Sahu2022,andrews2014bidirectional,brubaker2021optomechanical,delaney2021non,mirhosseini2020superconducting,forsch2020microwave,mirhosseini2020superconducting,bartholomew2020chip}. Neutral atoms are an attractive alternative transducer: they couple strongly to optical photons in their ground states, and to microwave/millimeter-wave photons in their Rydberg states~\cite{covey2019,petrosyan2019microwave,hafezi2012hyperfine}. Nonetheless, strong coupling of atoms to both types of photons, in a cryogenic environment to minimize thermal noise, has yet to be achieved. Here we demonstrate quantum-limited transduction of millimeter-wave (mmwave)~\cite{pechal_millimeter-wave_2017} photons into optical photons using cold $^{85}$Rb atoms as the transducer. We achieve this by coupling an ensemble of atoms simultaneously to a first-of-its-kind, optically accessible three-dimensional superconducting resonator~\cite{suleymanzade2020tunable}, and a vibration suppressed optical cavity, in a cryogenic ($5$~K) environment. We measure an internal conversion efficiency of $58(11)~\%$, a conversion bandwidth of $360(20)$~kHz and added thermal noise of $0.6$ photons, in agreement with a parameter-free theory. Extensions to this technique will allow near-unity efficiency transduction in both the mmwave and microwave regimes. More broadly, this state-of-the-art platform opens a new field of hybrid mmwave/optical quantum science~\cite{Schmiedmayer2015review,Clerk2020}, with prospects for operation deep in the strong coupling regime for efficient generation of metrologically or computationally useful entangled states~\cite{davis2016approaching} and quantum simulation/computation with strong nonlocal interactions~\cite{swingle2016measuring}.
	\end{abstract}
	\maketitle

Conversion of quantum information from stationary computation qubits to flying optical qubits lies at the heart of quantum networking tasks from linking remote quantum computers for scaling to fault-tolerance, to quantum key distribution~\cite{diamanti2016practical} and secondary technologies like quantum repeaters~\cite{briegel1998quantum}. In general, it is desirable for the transducer to (i) be efficient – the loss of information through various decay channels during the conversion process should be minimal; (ii) have a high bandwidth – the conversion should be much faster than the decay of the stationary qubits; (iii) be noiseless – the added noise, typically thermal, should be much smaller than one photon to preserve the integrity of the quantum information~\cite{svensson2020figures}.

There has been tremendous progress towards building such systems for superconducting qubits in recent years, particularly by using materials with direct electro-optic coupling~\cite{mckenna2020cryogenic,Xu2021,Sahu2022} or some combination of piezo-electric, electro-mechanical and opto-mechanical couplings~\cite{andrews2014bidirectional,brubaker2021optomechanical,delaney2021non,mirhosseini2020superconducting,forsch2020microwave}. In a recent landmark experiment, the state of a superconducting transmon qubit was transduced to the optical domain by coupling to a nanomechanical resonator that was also coupled to an optical mode~\cite{mirhosseini2020superconducting}. Although the added noise was low, the intrinsic conversion efficiency was limited to $10^{-3}$. In another experiment, a silicon nitride membrane, cooled near its mechanical ground state and coupled simultaneously to an LC resonator and an optical cavity, was used to efficiently read out the state of a qubit, with significant added noise and limited bandwidth~\cite{delaney2021non}.  These approaches have typically been limited by the adverse effects of optical photons on superconductors~\cite{barends2011minimizing} and added thermal noise from strong microwave and optical pumps~\cite{brubaker2021optomechanical}. The strong pumps are required not only to bridge the energy gap between microwave and optical domains, but also to achieve the strong couplings necessary for high efficiency transduction.
	\begin{figure*}
		\centering
		\includegraphics[trim = 0cm 15cm 0cm 0cm]{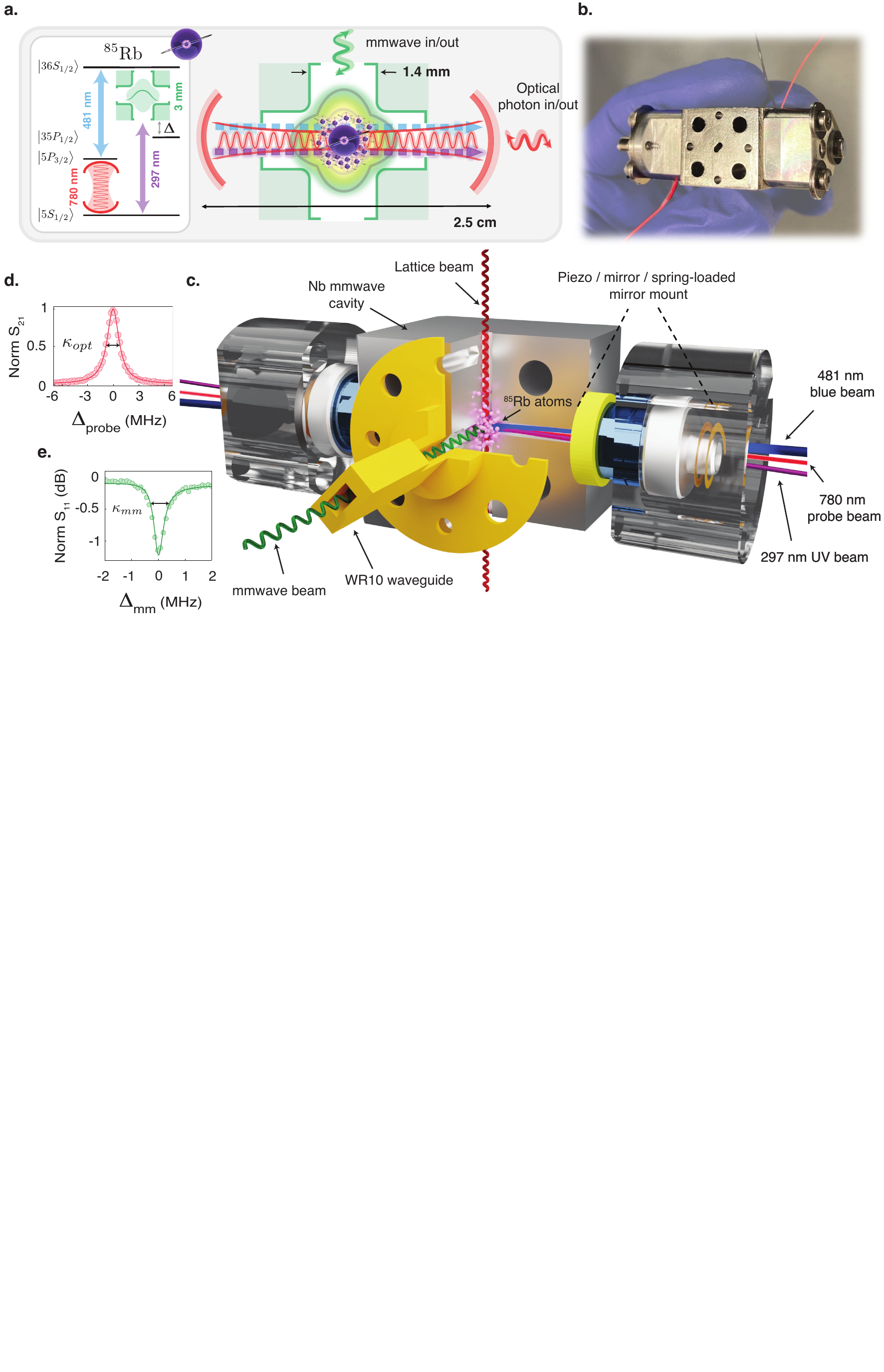} 
		\caption{
	    	\textbf{Experimental Setup.}
			\textbf{a,} Schematic of the system: Left - atomic energy levels and wavelengths of light involved in transduction, Right - the internal structure of the optical and mmwave interface. The atoms are coupled to an optical cavity (red) and a superconducting mmwave resonator (green) simultaneously. The interconversion loop is closed by the blue ($481$~nm) and UV ($297$~nm) control lasers.     
			\textbf{b,} Picture of the physical hybrid cavity.  
			\textbf{c,} Expanded view of the main assembly. The mmwave mode is hosted at the intersection of three orthogonal cylindrical waveguides machined in a niobium spacer. One of the waveguides is used to transport cold $^{85}$Rb atoms in a 1D optical lattice (red vertical beam) from a Magneto Optical Trap $6.5$~cm above the structure. A second waveguide provides optical access for forming the optical cavity by affixing mirrors and piezos at the two ends of the spacer using a spring-loaded design for vibrational stability (see SI~\ref{SI:optical cavity}). It also provides a path for the blue and UV lasers (shown in blue and purple respectively). The third waveguide allows us to probe the resonator with an external mmwave drive.
			\textbf{d,} Bare optical cavity transmission with full width half maximum (FWHM) linewidth $\kappa_{opt} = 2\pi\times 1.7$~MHz. The solid line is a Lorentzian fit.
			\textbf{e,} Reflection spectrum of the superconducting mmwave cavity at 5K with FWHM linewidth $\kappa_{mm} = 2\pi\times 800$~kHz. 
			\label{fig:setup}} 
	\end{figure*}

Neutral atoms offer a promising alternative to the aforementioned approaches~\cite{covey2019,petrosyan2019microwave,hafezi2012hyperfine}. In their ground states, atoms couple strongly to optical frequencies, and when excited to Rydberg states they have strong dipole transitions in the microwave regime. In fact, free-space interconversion with Rydberg atoms has been demonstrated at room temperature~\cite{vogt2019efficient,tu2022high}, but coupling the atoms to a high quality factor superconducting resonator at cryogenic temperatures is ultimately essential to enable transduction of a superconducting qubit with low thermal noise and high efficiency. Even though early experiments with transiting circular Rydberg atoms in high quality factor superconducting cavities laid the foundations of modern cavity and circuit quantum electrodyanmics~\cite{raimond2001review}, it has been challenging to combine those techniques with modern innovations in the control of neutral atoms~\cite{hermann2014long}.

In this work, we overcome these difficulties, strongly coupling cold Rydberg-dressed atoms to a superconducting millimeter wave resonator crossed with an optical cavity, and employ this unique platform to interconvert optical and millimeter wave (mmwave) photons. These breakthroughs are enabled by a novel 3D mmwave resonator and a vibration stabilized optical cavity in a tightly integrated design. Our mmwave resonator confines photons to a volume of $\approx \lambda^3/10$ while maintaining the optical access required to (i) form an optical cavity; (ii) load and (iii) optically address the atoms~\cite{suleymanzade2020tunable}. We individually characterize the coupling of the atoms to various fields involved in the transduction process by probing the transmission of the optical cavity, enabling us to construct a simple parameter-free model of interconversion. To measure the performance of our transducer, we drive the mmwave resonator and observe the output of the optical cavity and find it in good agreement with the model. We then compare the output of the optical cavity with and without mmwave drives and cleanly separate the interconversion of thermal photons from that of an applied coherent drive, both in the optical photon count rates and second order correlations, further confirming the ability to operate in the single photon regime.

Our mmwave resonator traps photons at the intersection of three orthogonal cylindrical waveguides drilled in a solid cuboid of high purity niobium (Fig.~\ref{fig:setup}b,d, and SI~\ref{SI:mmwave resonator}), while the waveguides themselves remain evanescent, preventing photon leakage~\cite{suleymanzade2020tunable}. We use one of the waveguides to make an optical cavity by mounting mirrors at the two ends. The mirrors and piezos are clamped using spring washers to minimize the effect of cryo-fridge vibrations (see SI~\ref{SI:optical cavity}). A second waveguide is used to load an ensemble of atoms in the center of the resonator using a one dimensional transport optical lattice. All optical fields necessary to address the atoms are routed through these two waveguides, with the Gaussian waists of the laser beams much smaller than the diameter of the waveguide. We use the third waveguide to couple in mmwave photons (see Fig.~\ref{fig:setup}d).

	\begin{figure*}[htb]
		\centering
		\includegraphics[trim = 0cm 18cm 0cm 0cm]{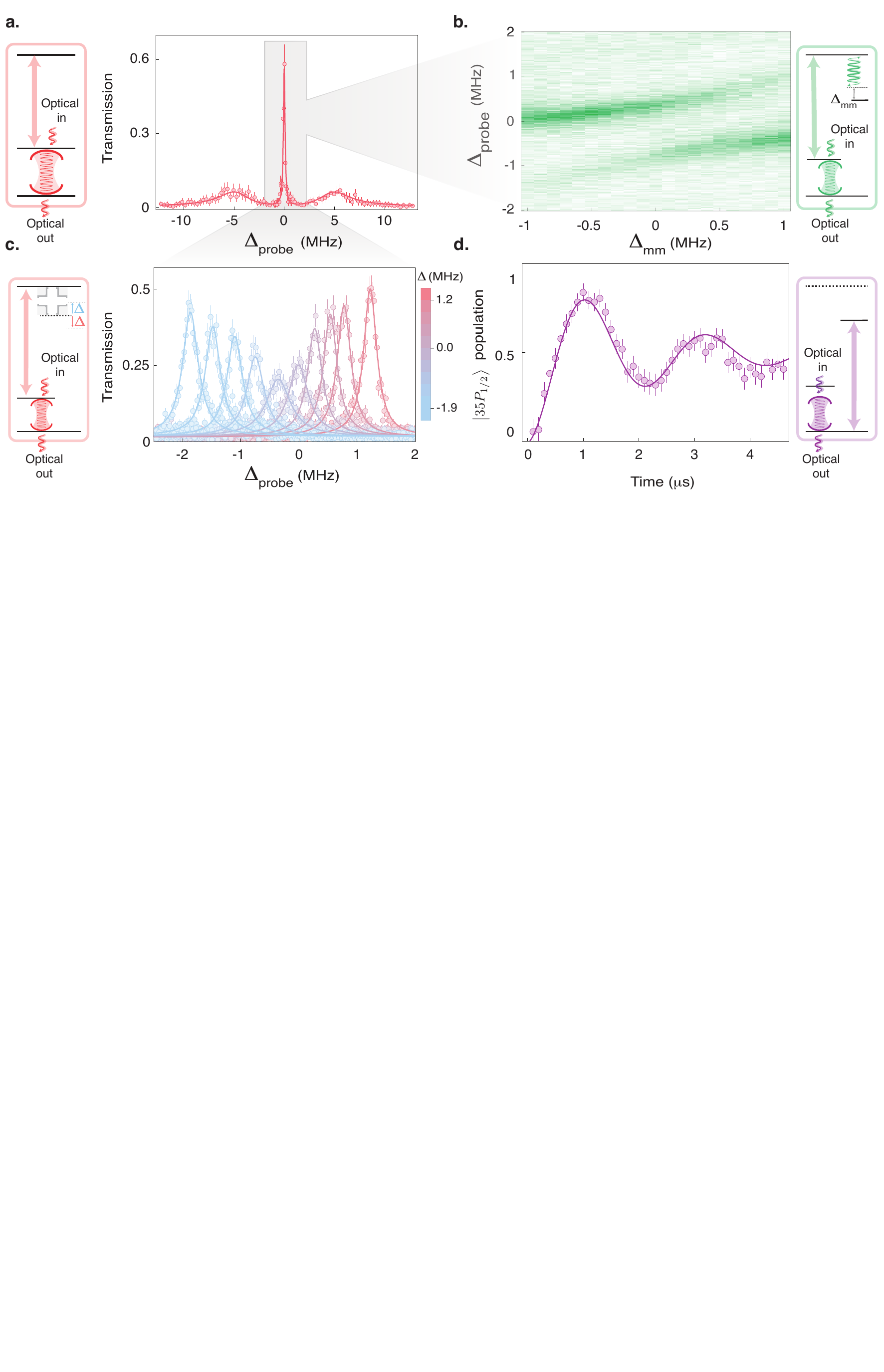}
		\caption{
			\textbf{The building blocks for transduction.} Each leg of the conversion loop can be independently characterized by probing the transmission of the optical cavity.
			\textbf{a,} A $780$~nm photon couples an ensemble of atoms to a collective $5\textrm{P}_{3/2}$ state, which is further coupled by a blue ($481$~nm) photon to a collective $36\textrm{S}_{1/2}$ Rydberg state. We observe the excitations of this system in the transmission of the optical cavity as we scan the $780$~nm probe laser across the cavity resonance (with detuning $\Delta_{\textrm{probe}}$). The atoms hybridize with the optical mode to form two bright polaritons (whose large width comes from the lossy $5\textrm{P}_{3/2}$ state $\Gamma\approx 2\pi \times 6$~MHz) and a dark polariton (narrow central feature), which only contains an excitation into the Rydberg state for its atomic component. We typically operate on resonance with the dark polariton for transduction, to prevent efficiency loss due to decay of the $5\textrm{P}_{3/2}$ state. \textbf{b,} Applying a coherent mmwave drive near the atomic $36\textrm{S}_{1/2} \leftrightarrow 35\textrm{P}_{1/2}$ resonance, but detuned by $\Delta\approx 2\pi\times 12.4$~MHz from the mmwave cavity splits the dark polariton peak due to admixing of the $36\textrm{S}_{1/2}$ and $35\textrm{P}_{1/2}$ Rydberg states by the mmwave drive. Scanning the mmwave drive frequency (with detuning $\Delta_{\textrm{mm}}$) produces an avoided crossing, enabling us to determine the $36\textrm{S}_{1/2}\leftrightarrow 35\textrm{P}_{1/2}$ resonance frequency, and therefore the detuning between this transition and the mmwave cavity. 
			\textbf{c,} To Stark-tune the atomic transition into resonance with the primary ``science" mmwave cavity mode we use an auxiliary, ``tuning", mmwave mode which is detuned from the atomic transition by $2\pi\times 1.9$~GHz. At atomic resonance with the mmwave science mode, we observe a significant broadening and reduction in peak height of the dark polariton transmission feature due to coupling with the vacuum- and thermal- fields of the mmwave cavity.
			\textbf{d,} Population in the $5\textrm{S}_{1/2}$ state after UV laser pulses of variable length. In the absence of the blue coupling beam and with the $36\textrm{S}_{1/2}\leftrightarrow 35\textrm{P}_{1/2}$ transition detuned from the mmwave cavity, we observe Rabi flopping which allows us to determine the UV Rabi frequency $\Omega_{uv} = 2\pi\times230$~kHz. We measure the ground vs. Rydberg atomic population via a dispersive shift of the optical cavity.
			\label{fig:syscharacterize}}
	\end{figure*}

Our transduction scheme may be understood as four-wave mixing mediated by an atomic ensemble as shown in Fig.~\ref{fig:setup}a. For the case of mmwave to optical conversion, the atoms are first off-resonantly coupled from the ground $5\textrm{S}_{1/2}$ to the $35\textrm{P}_{1/2}$ Rydberg state by a $297$~nm UV laser. Due to the detuning, the ensemble of atoms can resonantly \& collectively absorb a $297$~nm UV photon \emph{only} if it then also absorbs a signal mmwave photon (occupying mode $b$), leading to a collective excitation in the $36\textrm{S}_{1/2}$ state. The excited atom is then stimulated to emit a $481$~nm blue photon, de-exciting it to the $5\textrm{P}_{3/2}$ state and finally to spontaneously emit a $780$~nm signal photon into the optical cavity mode $a$, closing the loop back to the initial ground state and completing the transduction process. For the cavity emission to be collectively enhanced, it is vital for the loop to be closed to the same many-particle quantum state, including all atomic motional and internal degrees of freedom. This can also be understood as requiring that no information about the fields be left behind in the state of the atoms, which would destroy the interference that gives rise to collective enhancement. This enforces stringent phase and mode matching constraints that we satisfy by co-aligning the optical cavity mode with the blue and UV control fields. To avoid further complications from the magnetic substructure of the atoms, we optically pump them along the transport lattice direction in the presence of a large Zeeman field that splits the levels, and choose the polarizations of the optical fields such that the loop is deterministically closed (see Methods~\ref{SI:sequence} and~\ref{SI:optical pumping}). 

In the limit of a large detuning ($|\Delta|\gg|\Omega_{uv}|$) between the mmwave cavity and the $36\textrm{S}_{1/2}\leftrightarrow 35\textrm{P}_{1/2}$ atomic transition, the linearized interaction Hamiltonian (see SI~\ref{SI:interconversiontheory}) describing this system may be written as: 
\begin{equation}
    \mathcal{H}_{\textrm{int}}/\hbar = \sqrt{N} g_{opt} a E^{\dag} + \Omega_b E R^{\dag} + g_{mm}\frac{\sqrt{N} \Omega_{uv}}{\Delta} b^{\dag} R + h.c.
    \label{eqn:ham}
\end{equation}

where $E$ and $R$ are the bosonized collective excitation operators to the $5\textrm{P}_{3/2}$ and the $36\textrm{S}_{1/2}$ states respectively, from a reservoir of $N$ atoms in the $5\textrm{S}_{1/2}$ ground state, $g_{opt}$ and $g_{mm}$ are the single atom coupling strengths for the optical and mmwave modes, and $\Omega_b$ and $\Omega_{uv}$ are the Rabi frequencies for the blue and UV fields.

We now characterize the steps of the interconversion process by performing a series of experiments that take optical photons further and further around the transduction loop. We first omit the UV drive and detune the atoms from the mmwave cavity ($|\Delta|\gg|g_{mm}|$), leaving only 3-level atoms coupled to the optical cavity and blue control field (Fig.~\ref{fig:syscharacterize}a schematic) in the canonical Rydberg electromagnetically induced transparency (EIT) configuration~\cite{mohapatra2007coherent,ningyuan2016observation}. In this regime, the relevant optical excitations of the system are cavity polaritons~\cite{ningyuan2016observation}, quasiparticles consisting of a superposition of atomic and photonic components. The transmission of the optical cavity shows a dark polariton residing between two bright polaritons (Fig.~\ref{fig:syscharacterize}a). The dark polariton is narrow because its atomic part consists only of the long-lived Rydberg state, while the bright polaritons have a significant admixture of the short-lived $5\textrm{P}_{3/2}$ state. For interconversion, we therefore operate on the dark polariton resonance in order to minimize loss and required pump powers (see SI~\ref{SI:fwm}). In the limit that the dark polariton is spectrally resolved from the bright polaritons, the transduction process can be understood as a beam-splitter interaction between the mmwave and the dark polariton modes (see SI~\ref{SI:interconversion dark polariton }). Furthermore, fitting spectra like the one shown in Fig.~\ref{fig:syscharacterize}a to analytical models from non-hermitian perturbation theory~\cite{jia2018quantumdot} enables us to extract parameters that are central to building a quantitive model of interconversion: $N$, the atom number; $\Omega_b$ the blue Rabi frequency; and $\Gamma_R$, the collectively suppressed decoherence rate of the Rydberg state\cite{georgakopoulos2018theory} (see Methods ~\ref{SI:parameters} and Table~\ref{table:parameter table}).

Before introducing the quantized mmwave cavity field, we first study the impact of a classical mmwave field on the dark polariton resonance. This field drives the $36\textrm{S}_{1/2}$ atomic component of the polariton to the $35\textrm{P}_{1/2}$ state and leads to an Autler-Townes splitting of the dark polariton. The resulting avoided crossing as a function of mmwave drive frequency lets us extract the frequency of this atomic transition (Fig.~\ref{fig:syscharacterize}b) and thus precisely determine the detuning ($\Delta$) of the atoms from the mmwave cavity.

Thus prepared, we next explore the coupling of the atoms to the mmwave cavity vacuum by varying the detuning between atoms and the (undriven) mmwave cavity. In practice, this is achieved through an ac Stark shift of the atomic Rydberg levels by driving a far off-resonant (from atoms), ``tuning" mode of the mmwave cavity (see Methods~\ref{SI:tuning} and SI~\ref{SI:mmwave resonator}). We observe a significant broadening of the dark polariton when the atoms are resonant with the mmwave cavity (Fig.~\ref{fig:syscharacterize}c). This Purcell-like broadening is a result of vacuum induced coupling between the polariton and the $35\textrm{P}_{1/2}$ Rydberg state, enhanced by the non-zero thermal population of the mmwave cavity. The observed broadening is consistent with our calculated value of $g_{mm} = 2\pi\times182$~kHz and the expected thermal occupation $n_{th}\approx$ 0.6 photons for a $100$~GHz cavity at $5$~K (see Methods~\ref{SI:purcell}). 

The final ingredient required to close the interconversion loop is the direct UV ($297$~nm) transition between the ground state and the $35\textrm{P}_{1/2}$ state, which we explore directly, rather than investigating its impact on the dark polariton. We turn off the blue beam, return to a large detuning between the mmwave cavity and the atoms, and drive Rabi oscillations on the $5\textrm{S}_{1/2}\leftrightarrow 35\textrm{P}_{1/2}$ transition with UV pulses of varying lengths (see Fig.~\ref{fig:syscharacterize}d). The optical cavity is employed only to detect the transfer of atomic population via a dispersive shift of the cavity line~\cite{leroux2010squeezing}. This measurement directly calibrates $\Omega_{uv}$ (see Methods~\ref{SI:parameters} and Table~\ref{table:parameter table}), the final key parameter in Hamiltonian.

Before turning to transduction measurements, it is instructive to first understand the figures-of-merit that determine conversion efficiency. For the coupling scheme shown in Fig.~\ref{fig:setup}a, with the Hamiltonian given by Eqn.~\ref{eqn:ham}, it can be shown (see SI~\ref{SI:interconversiontheory}) that the conversion efficiency is given by:

\begin{equation}
\label{eqn:efficiency}
    \mathcal{E} = \frac{\kappa_{opt}^{ext}\kappa_{mm}^{ext}}{\kappa_{opt}\kappa_{mm}} \eta_{o\leftrightarrow mm}
\end{equation}

Here $\kappa_{opt}^{ext}$ and $\kappa_{mm}^{ext}$ are the outcoupling rates for the optical and mmwave cavities through the ports where they are driven and measured (vs absorption and other loss ports). Assuming single-ended cavities ($\kappa_j^{ext}=\kappa_j$), the conversion efficiency of the idealized optical $\leftrightarrow$ mmwave interconverter reduces to (see SI~\ref{SI:interconversiontheory}) $\eta_{o\leftrightarrow mm}=\frac{4 C_{o}C_b C_{mm}}{(C_b+(1+C_{mm})(1+C_o))^2}$, where $C_o=N_g\frac{4 g_{opt}^2}{\kappa_{opt}\Gamma}$ is the collectively-enhanced optical cooperativity, $C_{mm}=N_{mm}\frac{4 g_{mm}^2}{\kappa_{mm}\gamma_r}$ is the collectively-enhanced mmwave cooperativity, and $C_b=\frac{4\Omega_b^2}{\Gamma\gamma_r}$ is a generalized cooperativity defined for the blue EIT-control field. Here the number of effective ground state atoms in the presence of UV-dressing is $N_g=N\cos^2{\theta}$, while the number of effective atoms in the Rydberg P-state is $N_{mm}=N\sin^2{\theta}$ where $N$ is the total atom number, and $\tan\theta=\left|\frac{\Omega_{uv}}{\Delta}\right|$ controls the dressing fraction of the relevant UV-dressed ground state.

Conversion efficiency is optimized by ensuring good impedance matching in the coupling between optical and mmwave channels (see SI~\ref{SI:interconversiontheory}). At fixed $C_o$ and $C_{mm}$, $\eta_{o\leftrightarrow mm}$ may thus be maximized by varying $C_b$ (by changing $\Omega_b$), yielding $C_b=(1+C_o)(1+C_{mm})$, with $\eta_{max}=\frac{C_o C_{mm}}{(1+C_o)(1+C_{mm})}$. It is apparent that $\eta_{max}$ is then limited by the smaller of the two cooperativities. Our ability to \emph{divide} the collective enhancement between optical and mmwave transitions via the UV dressing fraction thus strongly relaxes the performance requirements of both resonators. In practice we are often limited by blue power, in which case we use the maximum available blue power and achieve impedance matching by varying $C_{mm}$ or $C_o$ via the detuning ($\Delta$) and atom number, and yielding $\eta_{max}\approx 1-\frac{2}{\sqrt{C_b}}$ (see SI~\ref{SI:interconversiontheory}). While this simple story is most accurate for a running wave $780$~nm cavity, residual effects from the non-phase-matched running wave in our standing wave cavity are suppressed by high cooperativity (see SI~\ref{SI:interconversiontheory}); all theory curves reflect a full standing wave model.

	\begin{figure}
		\centering
		\includegraphics[trim= 0.3cm 12.5cm 0cm 0cm]{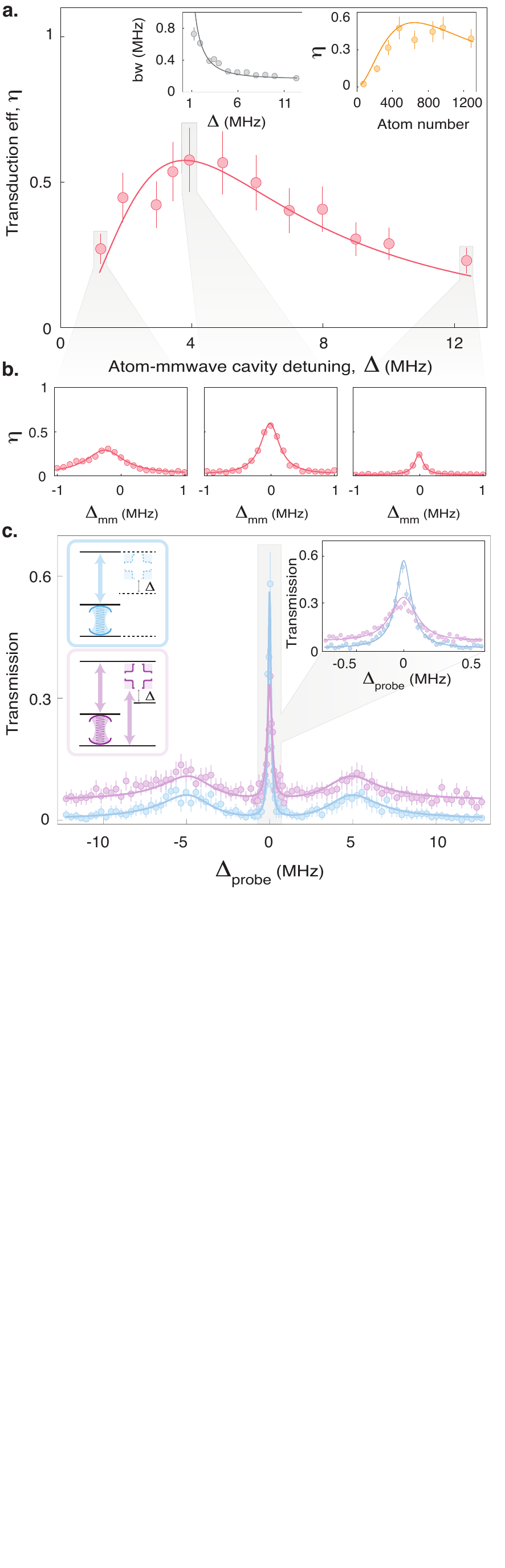} 
		\caption{
			\textbf{Transducer Performance.} 
	        \textbf{a,} Peak internal conversion efficiency (red) as the atom-mmwave cavity detuning ($\Delta$) is varied. The efficiency is maximized when the transduction process is impedance matched, which can also be controlled by varying the atom number, as in the second inset (yellow), with $\Delta = 2\pi\times 3.9$~MHz fixed.
			\textbf{b,} For each point in \textbf{a}, we scan the mmwave drive frequency (detuning from cavity $\Delta_{mm}$) through the science mode and fit the resulting spectrum to a Lorentzian with an offset to extract the peak conversion efficiency. We also extract the interconversion bandwidth (gray, first inset of \textbf{a}) from the FWHM of the fitted lorentzians. Here we operate with effective atom number $N\approx 600$.
		    \textbf{c,} We indirectly probe optical to mmwave transduction by observing the optical transmission with (purple) and without (blue) the $297$~nm UV laser ($\Delta\approx 2\pi\times4.2$~MHz and $N\approx500$). The inset zooms into the dark polariton, which is suppressed and broadened by the additional loss of conversion to mmwave photons. The offset reflects conversion of thermal mmwave photons into optical photons. All solid curves are parameter-free theory except the blue fit to the spectrum in \textbf{c}, which provides the parameters.
			\label{fig:interconversion}}
	\end{figure}

We are now prepared to combine all of the couplings and perform transduction. We measure $\eta_{o\leftrightarrow mm}$, the internal conversion efficiency, by exciting the mmwave cavity with a classical drive of known average intracavity photon number, $n_{ph}$, and measuring the converted photon count rate at the output of one of the ports of the optical cavity (see Methods~\ref{SI:conversion efficiency}). For our system, it is always favorable to fix the blue and UV powers to their maximum achievable values. We typically operate at $n_{ph} \approx 2$.

\begin{figure*}[htb]
\centering
\includegraphics[trim = 0cm 23cm 0cm 0cm]{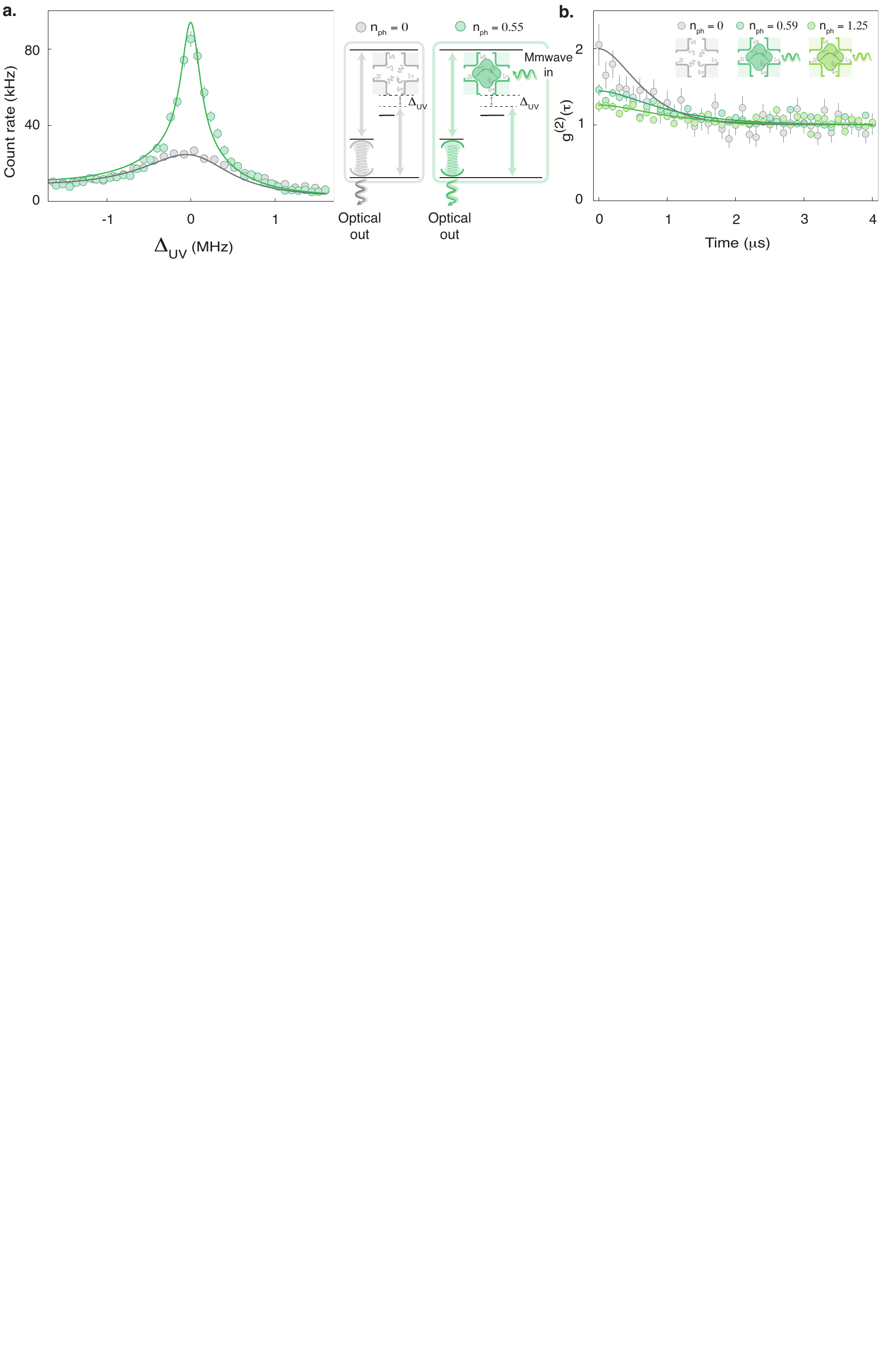}
\caption{\textbf{Thermal photons and intensity correlations.}
	\textbf{a,} Count rate at the output of the optical cavity without an external mmwave drive (gray) and with a coherent mmwave drive with approximately 0.5 photons (green), as the frequency of the UV laser is scanned. The broad gray feature reveals the interconversion of the broadband thermal drive of the mmwave cavity, while the conversion of a coherent drive quickly becomes lossy as the UV laser is tuned away from the dark polariton resonance. 
	\textbf{b} Measured second-order correlation functions of converted optical photons for various mmwave drive strengths : 0 drive photons (gray), 0.5 photons (dark green) and 1.1 photons(light green). For both \textbf{a} and \textbf{b}, the solid curves are theory without any free parameters. The statistical error in $n_{ph}$ is $\sim\pm 10~\%$.
	\label{fig:coherences}}
	\end{figure*}
	
To systematically study interconversion, we first vary $\Delta$, which controls $C_{mm}$ through the steady state dressing fraction in the $35\textrm{P}_{1/2}$ state. A clear maximum in the conversion efficiency can be seen near $\Delta \approx 2\pi\times 4$~MHz, where the transduction process is impedance matched (Fig.~\ref{fig:interconversion}a). If we instead fix the detuning and vary the atom number (changing both $C_o$ and $C_{mm}$), we observe a similar impedance matching in the variation of the conversion efficiency (Fig.~\ref{fig:interconversion}a inset). For each of these efficiency measurements, we also measure the interconversion bandwidth (Fig.~\ref{fig:interconversion}a inset, and Fig.~\ref{fig:interconversion}b), which directly probes $|S(\omega)|^2$, where $S(\omega)$ is the transfer function of the interconverter (see SI~\ref{SI:interconversiontheory}). At its highest, we measure an efficiency of $58(11)~\%$, with a bandwidth of $360(20)$~kHz, all in agreement with our parameter-free theory (solid curves).

Due to the mmwave attenuation built into our system to limit room-temperature black-body heating of the mmwave cavity (see Fig.~\ref{fig:SupFig1}), we cannot directly probe the reverse, optical-to-mmwave conversion process. We instead observe it indirectly through the loss of optical photons when measuring the optical cavity transmission. To achieve this, we operate without an external mmwave drive and compare the optical transmission with and without the UV beam, measuring a reduction and broadening of the height of the dark polariton resonance (Fig.~\ref{fig:interconversion}c). The lost optical photons correspond not only to interconverted mmwave photons, but also photons lost to free space scattering, and additional reflection of the optical field off the optical cavity due to changes in impedance matching. Nonetheless, we find that our measurement is in good agreement with our theory and the predicted conversion efficiency. Additionally, the optical cavity transmission in the presence of the UV beam experiences a probe-frequency-independent background due to transduced thermal mmwave cavity photons.

This background due to converted thermal photons is also present as an offset in Fig.~\ref{fig:interconversion}a and b, but it is less apparent in the presence of a coherent drive. To characterize the thermal noise, we operate the transducer near the nominally impedance matched point in Fig.~\ref{fig:interconversion}a, vary the UV laser frequency to translate the transduction band in frequency, and plot the observed count rate at the one of the optical cavity ports (Fig.~\ref{fig:coherences}a). Without an external mmwave drive, we observe a broad asymmetric feature which develops a narrow peak as a weak coherent drive ($n_{ph} = 0.55(5)$) is added. The narrow feature is a direct measurement of the transduction bandwidth, while the broader feature reflects the fact that the thermal backgrounds have a bandwidth given by the mmwave cavity, which is larger than that of the transducer. The asymmetry arises because the impedance matching changes significantly as the UV is tuned towards the atomic resonance (lower $\Delta_{\textrm{UV}}$ in Fig.~\ref{fig:coherences}a). Because many of the mmwave cavity thermal photons are outside of the transduction band (due to the larger bandwidth of the mmwave cavity), the measured count rate for thermal photons is significantly lower than it would be for a coherent drive with the same average photon number. The shape of the thermal feature as well as the peak count rates are well matched by predictions from the Hamiltonian in Eqn.~\ref{eqn:ham} with the $0.6$ thermal photons expected at $5$~K (see Methods~\ref{SI:theory calc}). This also implies that the added noise, when referred to the input, is simply $0.6$~photons~$\textrm{sec}^{-1}\textrm{Hz}^{-1}$ -- the power spectral density of the input thermal drive at the frequency of the mmwave cavity.

To ensure that the observed noise photons are indeed thermal and not coherent backgrounds (for example from the off-resonant tuning drive) we measure the second order intensity correlation function of the converted optical photons $g^{(2)}(\tau)$ as we vary the external coherent mmwave drive strength (Fig.~\ref{fig:coherences}b). In the absence of the coherent drive, $g^{(2)}(0)\approx 2$ as expected for a thermal source. Once the coherently injected photon population exceeds the thermal population, $g^{(2)}(0)$ drops towards 1.
Even for an entirely thermal state, the decay time of $g^{(2)}(\tau)$ is not set solely by the mmwave cavity linewidth, but rather by the (lower) bandwidth of the transducer. The calculated $g^{(2)}(\tau)$ (see Methods~\ref{SI:theory calc}) for $0.6$ thermal photons (a temperature of $5$~K) is in good agreement with the data.

These experiments demonstrate quantum limited transduction between mmwave and optical photons with a high efficiency and large bandwidth.
We envision clear paths to further improving the transducer performance and coupling to superconducting qubits: The conversion efficiency in our transducer is presently limited by the Rabi frequency of the blue laser. Modifications to the mmwave mode and blue/UV laser polarizations will significantly increase the blue, UV and the mmwave resonator coupling strengths and thereby push the efficiency to near unity (see Methods~\ref{SI:improvements}). At present, the added noise is set by the $5$~K temperature of the mmwave cavity structure, so cooling down to $1$~K would reduce this noise from $0.6$ photons to $<0.01$ photons and significantly improve the mmwave cavity quality factor. Direct coupling between atoms and ($\sim 10$~GHz) transmon qubits will be possible by working at higher principal quantum number atomic states in larger, colder resonators~\footnote{The reduced atom-resonator coupling due to increased mode volume may be offset by larger P -Rydberg state dressing fraction; Rydberg-Rydberg interactions will remain negligible due to the diluteness of the atomic cloud.}. Introducing a high-kinetic inductance resonator~\cite{anferov2020millimeter,pechal_millimeter-wave_2017} as a mmwave-to-microwave interface may even obviate the need for direct microwave transitions in the Rydberg atoms.

More broadly, this work paves the way to strong coupling between single atoms and mmwave resonators while retaining the optical access that enabled recent advances in single atom control and detection~\cite{endres2016atom}. Owing to the small mode volume of the mmwave resonator and the large dipole moment of the atoms in the Rydberg states, our system can achieve single particle cooperativities as high as $6000$ at $\approx 2$~K temperatures~\cite{suleymanzade2020tunable}. The resulting strong all-to-all interactions between atoms open a new parameter regime for generating squeezed~\cite{clerk2022squeezing} and other highly entangled spin states~\cite{davis2016approaching} for metrology, as well as quantum simulation of exotic non-local physics including scrambling~\cite{swingle2016measuring}. Adding state-of-the-art single atom control to our setup would enable neutral atom quantum processors with global connectivity for entangling gates.

	\droptocpage
	
	\section{Acknowledgements}
    Funding for this research was provided by the National Science Foundation (NSF) through QLCI-HQAN grant 2016136, the Army Research Office through MURI grant W911NF2010136, and the Air Force Office of Scientific Research through MURI grant FA9550-16-1-0323. It was also supported by the University of Chicago Materials Research Science and Engineering Center, which is funded by the NSF under award DMR-1420709. M.S. acknowledges support from the NSF GRFP. We acknowledge Professor Erling Riis and Dr. Paul Griffin for fabrication of the grating used for our cryogenic MOT.

	\section{Author Contributions}
	The experiments were designed by A.K., A.S., M.S., L.T., D.S. and J.S. The apparatus was built by A.K., A.S., M.S., and L.T. The collection of data was handled by A.K and L.T. All authors analyzed the data and contributed to the manuscript.\\
	These authors contributed equally: A.K., A.S. and M.S.
    
	\section{Author Information}
	The authors declare no competing financial interests. Correspondence and requests for materials should be addressed to J.S. (jonsimon@stanford.edu). 
	
	\section{Data Availability}
	Due to the proprietary format of the experimental data as collected for this manuscript, it is available from the corresponding author upon request.
	
	\bibliographystyle{naturemag}
	\bibliography{ManuscriptBib}

\renewcommand{\appendixname}{Methods}
	
\setcounter{equation}{0}
\setcounter{figure}{0}
\renewcommand{\theequation}{M\arabic{equation}}
\renewcommand{\thefigure}{M\arabic{figure}}
\renewcommand{\thetable}{M\arabic{table}}

\clearpage

\setcounter{secnumdepth}{2}

\section*{Methods}

\subsection{Experimental Sequence}
\label{SI:sequence}

Our experiments take place in the cryogenic vacuum chamber shown in Fig.~\ref{fig:SupFig1}c. They begin with a MOT of laser-cooled $^{85}$Rb atoms loaded from a Rb dispenser. The dispenser is contained in a room-temperature box (see Fig.~\ref{fig:SupFig1}a and SI~\ref{SI:apparatus}) to enable the atoms to thermalize to $300$~K from their emission temperature of $750$~K. The thermalized atoms leak through a hole in the MOT grating into the trapping region, where they are cooled and trapped at the zero of a quadrupole field, using lasers on the Rb D$_2$ line at $780$~nm. $^{85}$Rb is chosen, instead of the more typical $^{87}$Rb, because of its higher isotopic abundance.

The MOT is loaded into a $785$~nm transport lattice via polarization gradient cooling to $5\mu$K (see SI~\ref{SI:apparatus}). The atoms are then transported from the MOT region into the crossed optical and mmwave cavity science region, in $14$~ms, by smoothly detuning one of the lattice beams using an RFSoC-based scriptable DDS based upon custom firmware, that drives double-passed AOMs.

Once the atoms are in the science region, they are optically pumped by a beam propagating along the transport lattice transport direction, into $\ket{\textrm{F}_g=3,\textrm{m}_F=3}$ state, using $\sigma^+$ polarized light on the the $\textrm{F}_g=3\leftrightarrow \textrm{F}_e=3$ transition of the D2 line. A magnetic field (about $3.2$~G) required for optical pumping was previously ``frozen in" to the superconducting cavity along the lattice direction (see SI~\ref{SI:optical pumping}). We then reduce the lattice depth to about $20$~\%, or turn the lattice off entirely to avoid broadening of the dark polariton due to different light shifts for different atoms depending on their position in the lattice.

A typical experiment involves observing the output of the optical cavity on one of the ports while (i) probing with $780$~nm light on the other port, (ii) while operating the transducer, or (iii) both as in Fig.~\ref{fig:interconversion}c. In experiments involving the UV beam, we limit the cumulative ``on-time" of the UV beam to $<300$~$\mu$S per shot. This is to avoid optical damage to the mirror coating from UV exposure. When external mmwave drives are required (either as signal photons for conversion or for off-resonantly tuning the atomic transition), we turn them on when the atom transport is initiated. The transduction process is controlled by turning on or off the blue and the UV beams. For experiments in Fig.~\ref{fig:interconversion}a and b, each shot typically involved five $10$~$\mu$S long transduction steps separated by $50$~$\mu$S of optical pumping. The optical pumping is required because the transduction process leads to depolarizion of atoms primarily due to the decay of atomic population from the $5\textrm{P}_{3/2}$ state, since our optical cavity is not on a cycling transition (see Methods~\ref{SI:optical pumping} and Fig.~\ref{fig:sup levels}). The rate of this depumping depends on the mmwave drive strength and repeated optical pumping is not always necessary, as is the case of measuring correlations of thermal photons in Fig.~\ref{fig:coherences}, which just involved a single $200$~$\mu$S long transduction step in each shot. All the optical signals are collected with two single mode fiber coupled single photon counting modules (Excelitas SPCM AQRH-14-FC) after a non polarizing beam splitter; the TTL pulses generated by the individual photon arrivals are time-tagged with $\sim 10$ns resolution, and stored for later analysis, using custom firmware in an FPGA.

Directly driving and probing the mmwave cavity is achieved with the $100$~GHz circuit shown schematically in Fig.~\ref{fig:SupFig1}b. The ``science" mode of the cavity is at $\approx 99.424$~GHz and the ``tuning" mode is at $\approx 101.318$~GHz. For transduction experiments, the \emph{science} mode power is actively stablized by diverting most of it to a Mi-wave 950W/387 mmwave power detector and feedback to a Mi-wave 900WF-30/388 voltage controlled attenuator (VCA).

\subsection{Polarizations, freezing magnetic fields and optical pumping}
\label{SI:optical pumping}

The detailed level scheme and various light polarizations for our experiment are shown in Fig.~\ref{fig:sup levels}a. The mmwave resonator was designed and tuned such that the ``science" mode is polarized along the optical cavity axis. While this was chosen, in part, to simplify couplings when atoms are optically pumped along the optical cavity axis (for experiments to explore non-linear physics), it turned out to be detrimental for transduction. For experiments in this paper, the optical pumping and therefore the quantization axis are chosen to lie along the transport lattice direction. Two main constraints dictated by transduction led to this choice. The first constraint is phase matching of the transduction fields (see SI~\ref{SI:interconversiontheory} and SI~\ref{SI:qpm}) which demand that the $k$ vectors of the absorbed (emitted) UV and mm-wave fields be equal to the k vectors of the emitted (absorbed) probe and UV fields:
\begin{equation}
    \vec{k}_{UV}+\vec{k}_{mm}=\vec{k}_{probe}+\vec{k}_{blue}
\end{equation}
Since $1/\vert \vec{k}_{mm} \vert$ is much larger than the size of the atom cloud, the effects of the phase variation of the mmwave mode can be neglected. Thus the phase matching condition requires the blue and the UV laser beams to have the same propagation direction as the cavity probe. The second constraint is that the net polarization change along the conversion loop should be 0, such that the loop begins and ends in the same atomic total angular momentum state.

The level scheme and polarizations shown in Fig.~\ref{fig:sup levels}a is one of the ways to satisfy these constraints along with the limitation that the mmwave mode is polarized along the optical cavity axis. The $780$~nm probe (or interconverted $780$~nm photons) and the blue are $\pi$-polarized. While sending in pure $\sigma^-$ polarized UV field would have been ideal for $\Omega_{uv}$, the phase matching constraint restricts us to send in linearly polarized light, out of which the $\sigma^-$ component couples to the ground state $5\textrm{S}_{1/2}\leftrightarrow 35\textrm{P}_{1/2}$ transition and $\sigma^+$ does not couple to the ground state at all. Similarly, while the mmwave cavity field is linearly polarized for this choice of quantization axis, only the $\sigma^+$ component couples to the $35\textrm{P}_{1/2}$, $m_j=-1/2$ state. 

The superconducting niobium (Nb) spacer shields the location of atoms at the intersection of the waveguides from any externally applied in magnetic fields not applied prior to cool-down. To impose the optimum magnetic fields for optical pumping, the spacer is heated to about 10 K (above superconducting temperature of Nb $\sim 9.2$~K) using a heating resistor thermally shorted to the spacer. 

At 10~K, applied B-fields are not repelled by the Meissner effect, so we optimize the optical pumping beam polarization and the required magnetic fields by measuring the depumping to the $F=2$ state from optical pumping beam (without a repump) through the vacuum Rabi splitting of the optical cavity transmission. One subtle point is that we have to wait about $300$~ms for the fields to settle after applying them due to Eddy currents. This is necessary because the subsequent ``freezing" of the fields is done in steady state. Once the correct fields are determined, we hold them and cool the spacer to $5$K. This freezes the field inside the cylindrical waveguides due to appearance of persistent currents -- a combination of Lenz's law and Meissner effect~\cite{meissner1933newer}.

Since these fields will stay frozen inside the cavity throughout, they also Zeeman-shift the $35\textrm{P}_{1/2}$ and $36\textrm{S}_{1/2}$ levels during probing. The choice of the magnitude and sign of the frozen field is dictated by the desired detuning between the $35\textrm{P}_{1/2}\leftrightarrow 36\textrm{S}_{1/2}$ mmwave transition and the mmwave cavity -- it should be in a range such that ac Stark shifts from classical fields applied to the ``tuning'' mode of the mmwave cavity can shift the atoms to resonance with the ``science" mode, but far enough so as to not induce Purcell broadening without tuning. For this experiment, the field was chosen such that the transition was $\sim12.4$~MHz detuned from the mm-wave cavity.

\subsection{Stark tuning the mmwave transition}
\label{SI:tuning}

We coarse tuned the mmwave cavity close to the atomic transition by using a combination of etching and mechanical squeezing, resulting in a cavity transition frequency of $99.42376$~GHz, roughly $6$~MHz from atomic transition without a magnetic field (see SI~\ref{SI:mmwave resonator} for details). While in principle it is possible to use magnetic fields to bridge this gap, it is desirable to have more ``real-time" control (which does not require re-freezing fields) over this detuning to thoroughly explore the parameter space. 

We achieve this by generating a mmwave Stark shift for both the $36\textrm{S}_{1/2}$ and $35\textrm{P}_{1/2}$ states using the field of the tuning mode at $101.318$~GHz, as shown in Fig.~\ref{fig:Mmcavmodes} and Fig.~\ref{fig:sup tuning}. In practice, we vary the tuning mmwave field power and measure the stark shift of the EIT line (and thus the $36\textrm{S}_{1/2}$ state). We use the calibration shown in Fig.~\ref{fig:sup tuning}b to extract the $35\textrm{P}_{1/2}$ state shift, and thus the atomic resonance frequency shift. We exactly diagonalize a Hamiltonian containing all magnetic sublevels of the $36\textrm{S}$, $35\textrm{P}$, and $35\textrm{S}$ manifolds and find the calculation to be in excellent agreement for the observed relative shifts of our $36\textrm{S}_{1/2}$ and $35\textrm{P}_{1/2}$ states.

\subsection{Purcell-like broadening and the non-linear Hamiltonian}
\label{SI:purcell}
Without the UV beam, our system is described by the non-linear Hamiltonian (derived in SI~\ref{SI:interconversiontheory}):

\begin{multline}
\label{eqn:non lin ham}
H=\delta_{a}\hat{a}^{\dagger}\hat{a}+\delta_{b}\hat{b}^{\dagger}\hat{b}+\delta_{e} E^{\dag} E+\delta_{r} R^{\dag} R+\Delta F^{\dag} F \\
+\left(g_{opt}\sqrt{N}\hat{a} E^{\dag}+h.c.\right)+\left(\Omega_{b} E R^{\dag}+h.c.\right)\\+(g_{mm} \hat{b}^{\dag} F^{\dag}R+h.c.)
\end{multline}

where in addition to the collective operators defined in the main text, we have introduced a new operator, $F^\dag$, which creates a collective excitation in the $35\textrm{P}_{1/2}$ state. The last term in the Hamiltonian~\ref{eqn:non lin ham} captures the exchange of a collective excitation between the $35\textrm{P}_{1/2}$ and $35\textrm{S}_{1/2}$ states through the absorption or emission of a mmwave photon. It describes a Jaynes-Cummings like coupling that makes the dynamics non-linear. Although this non-linearity is quite weak in our system due to the large mmwave cavity linewidth (resulting in a single particle cooperativity of around unity), we can observe its signature in the Purcell-like broadening of the data in Fig.~\ref{fig:syscharacterize}c.

\begin{figure*}
\centering
  \includegraphics[width=1\linewidth]{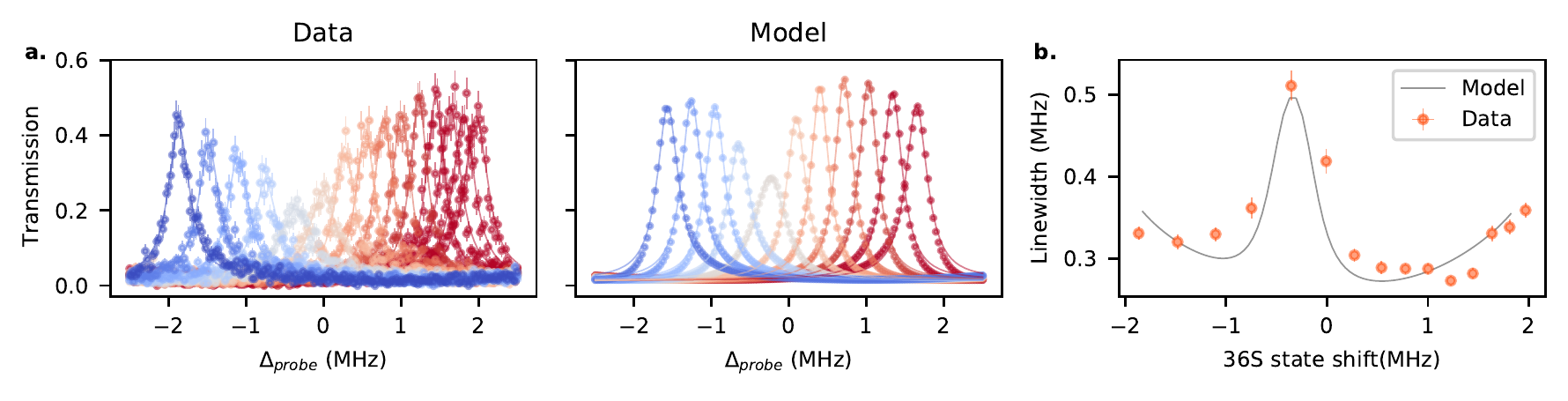}
\caption{\textbf{Purcell-like broadening of the dark polariton}
\textbf{a,} Same data as Fig.~\ref{fig:syscharacterize}c, but the full data set with spectra at additional atom-mmwave cavity detunings ($\Delta$ in main text), which were omitted from the plot in the main text for clarity. For this dataset, we simply varied the strength of the ``tuning" field to shift the atomic states, but did not change the blue frequency -- which would be required to keep the dark polariton at the same frequency as the optical cavity (at $\Delta_{probe}=0$ MHz). The asymmetry arises because at the point which the mmwave atomic transition is resonant with the mmwave cavity, the \emph{optical cavity} and blue beam are not resonant with the $5\textrm{S}_{1/2}\leftrightarrow36\textrm{S}_{1/2}$ transition.
\textbf{b,} Master equation simulation of the data in \textbf{a}, using the non-linear Hamiltonian derived in SI~\ref{SI:interconversiontheory} (equation~\ref{eqn:collective ham-no uv}), where all the parameters were experimentally measured using spectra with and without the blue beam at a large $\Delta$, except $g_{mm}$ and the number of thermal photons, which were calculated from first principles. In both \textbf{a} and \textbf{b}, the solid lines are Lorentzian fits.
\textbf{c,} The linewidths obtained from the Lorentzian fits of both theory and experiment. We find excellent agreement between the two, building confidence in our model. The dark polariton starts to broaden again near the edges because of mixing with the bright polaritons.
    }
\label{fig:sup purcell}
\end{figure*}

As we describe later in Methods~\ref{SI:conversion efficiency}, we employ this model to calibrate dispersive shifts of the dark polariton/EIT line to the coherently driven photon number in the cavity ($n_{ph}$) and show that a n\"aive calculation based on lowest order Stark shift would lead us to significantly overestimate conversion efficiency. To explore the validity of the model, we numerically solve the master equation for this Hamiltonian with most parameters independently measured by fitting EIT and vacuum Rabi splitting (VRS) spectra at a large detuning ($\Delta = 2\pi\times12.4$~MHz), where the non-linearity is inconsequential. We calculated (see Methods~\ref{SI:parameters}) $g_{mm}$ and the number of thermal photons from first principles. The results are shown in Fig.~\ref{fig:sup purcell} and we find good agreement between data and model.

\subsection{Relevant parameters for transduction}
\label{SI:parameters}
Table~\ref{table:parameter table} summarizes all the experimentally measured or theoretically calculated parameters for our system. The key parameters that affect transduction are the ones that determine the three cooperativites mentioned in the main text.

On the optical transition, we measure $\kappa_{opt}$ by probing the transmission of the bare optical cavity and fitting it to a Lorentzian (Fig.~\ref{fig:setup}c). We calculate the single atom-optical cavity coupling, $g_{opt}$ from the known length of the optical cavity, the radii of curvature of the cavity mirrors and dipole matrix element for the $\ket{5\textrm{S}_{1/2},F = 3,m_F = 3} \leftrightarrow \ket{5\textrm{P}_{3/2},F = 4,m_F = 3}$ atomic transition~\cite{tanji2011interaction}. Knowing $g_{opt}$, the effective atom number $N$ can be deduced by measuring the Vacuum Rabi Splitting, $2\sqrt{N}g_{opt}$, from the transmission spectrum of the optical cavity without the blue and the UV beams. For the blue transition, we measure the blue Rabi frequency, $\Omega_b$ and the decoherence rate of the collective $36\textrm{S}_{1/2}$ Rydberg state, $\Gamma_R$ by fitting the EIT spectrum to analytical results from non hermitian perturbation theory. These methods have been previously detailed in~\cite{jia2018quantumdot}. We note that while our atom cloud temperature of $4$~$\mu$K suggests a Doppler decoherence rate of $2\pi \times 150$~kHz, the presence of the blue beam suppresses this decoherence~\cite{georgakopoulos2018theory}.

On the mmwave side, calculating the single atom-mmwave cavity coupling, $g_{mm}$, requires the knowledge of the electric field profile in the mmwave cavity mode and the dipole matrix element of the $\ket{35\textrm{P}_{1/2},m_J = -1/2} \leftrightarrow \ket{36\textrm{S}_{1/2},m_J = 1/2}$ transition. We simulate the electric field profile using the finite element method (in the software package Ansys HFSS). The dipole matrix element for the atomic transition is calculated using the Atomic Rydberg Calculator (ARC) python package~\cite{ARC}. We typically determine $\kappa_{mm}$ from the reflection spectrum of the mmwave cavity using the circuit shown in $Fig.~\ref{fig:SupFig1}$. We can also measure it ``in-situ" using the atoms when they are far detuned ($2\pi\times12.4$~MHz) from the cavity, by scanning the mmwave drive frequency across the cavity resonance and plotting the ac-Stark shift of the dark polariton (see Fig.~\ref{fig:sup levels}c). We find the two values to be in good agreement, showing that the presence of the high powered blue laser does not significantly affect the mmwave cavity.

We note here that the decoherence rate of the $35\textrm{P}_{1/2}$ state, $\Gamma_F$, is inconsequential for the transduction process in our operation regime. This is true as long as $|\Delta|\gg\Gamma_F/2$, where the $|\Delta|$ is the magnitude of the detuning between the atoms and the mmwave cavity, or equivalently, between the atoms and UV beam in case of resonant operation. This can be understood in the following way: the UV beam drives atoms into a coherent spin state at a rate that is set by $\Delta$, and as long as this is much faster than $\Gamma_F$ (which for us is set by Doppler broadening), the coherence is continuously refreshed.

\subsection{Measuring internal conversion efficiency}
\label{SI:conversion efficiency}
We measure $\eta_{o\leftrightarrow mm}$, the internal conversion efficiency, by driving the mmwave cavity and measuring the photon count rate at the output of one of the ports of the optical cavity. It can be shown (see SI~\ref{SI:interconversiontheory}) that $\eta_{o\leftrightarrow mm} = \frac{1}{f_o}\frac{\kappa_{opt} }{\kappa_{opt}^{ext}}\frac{4 (R-R_{th})}{\kappa_{mm} n_{ph}}$, where $R$ is the count rate measured at the SPCMs with a coherent drive, $R_{th}$ is the background rate-primarily set by the interconversion of thermal mmwave photons, $f_o$ is the efficiency of the optical the path outside the optical cavity including the photon detection efficiency of the SPCMs, $\kappa_{opt}^{ext}$ is the out-coupling rate of the relevant port of the optical cavity and $n_{ph}$ is the expected photon occupation of just the bare mmwave cavity for a given coherent drive strength (minus the thermal occupation). We measure the efficiency of the optical path excluding the SPCM efficiency to be $\approx 0.5$ and independently, the efficiency of the SPCMs to be $\approx 0.56(5)$. We measure $\frac{\kappa_{opt}^{ext}}{\kappa_{opt} } = 0.63$ by reflection spectroscopy on the relevant optical port (see Fig.~\ref{fig:sup levels}b). This measurement as well as the optical cavity linewidth are in line with the measured mirror reflectivities from the manufacturer (Layertec GmbH).

We calibrate $n_{ph}$ by probing the dark polariton resonance weakly (in absence of the UV) and measuring its shift as the mmwave cavity is driven in the dispersive limit (see Fig.~\ref{fig:sup dispersive}a) , $\Delta = 2\pi \times 1.4$ MHz $\gg g_{mm} = 2\pi \times 180$ kHz. To the lowest order in perturbation theory, the shift of the the EIT peak is $\approx \frac{n_{ph} g_{mm}^2 sin^2(\theta_D)}{\Delta}$, where $\theta_D = tan^{-1}\left|{\frac{N g_{opt}}{\Omega_b}}\right|$ is the dark state rotation angle. Intuitively, this is just the lowest order Stark shift of the atomic component of the dark polariton. By comparing to master equation simulations of the non-linear Hamiltonian~\ref{eqn:non lin ham}, we actually find that this expression underestimates the the photon occupation and overestimates the conversion efficiency by about $15\%$; going to one order higher in perturbation theory yields a modified analytical expression that matches well with the master equation prediction (see Fig.~\ref{fig:sup dispersive}b and SI~\ref{SI:dispersive shift}). We therefore use the modified expression to calibrate $n_{ph}$. The lowest $n_{ph}$ we typically calibrate with this method is about 2 photons. While we can resolve the smaller shifts induced by weaker drives, to avoid the effect of systematic fluctuations and to get a much stronger signal, we calibrate weaker drives (as in Fig.~\ref{fig:coherences}) by instead measuring the count rate of interconverted photons and comparing it to the count rate of an dark polariton-shift-calibrated stronger drive. This assumes linearity of the transduction process, but both our measurements and mean-field simulations indicate that for few photon drives, the process is comfortably within the linear regime.

\subsection{Calculating optical photon count rates and correlation functions}
\label{SI:theory calc}
For the linearized system, it can be shown that

\begin{equation}
    a_{out}(\omega) = S(\omega)b_{in}(\omega)
\end{equation}

where $a_{out}(\omega)$ is the output optical field, $b_{in}(\omega)$ is the mmwave input field, and $S(\omega)$ is the transfer function of the transducer between the external coupling port of the mmwave cavity and the measurement port of the optical cavity. The details of calculating $S(\omega)$ for our system are covered in SI~\ref{SI:interconversiontheory}. We are interested in correlation functions of the form:
\begin{equation}
    \overline{g}^{(1)}(\tau) = \langle a_{out}^{\dag}(t+\tau)a_{out}(\tau) \rangle
\end{equation}
The rate of the transduced optical photons at the output port of optical cavity is simply given by $\overline{g}^{(1)}(0)$.
It can be shown (see SI~\ref{SI:interconversiontheory}) that for a purely coherent input mmwave field of strength $\beta$ 
\begin{equation}
    \overline{g}_{coh}^{(1)}(\tau) = \beta^2 |S(\omega_D)|^2 e^{-i \omega_{D} \tau}
\end{equation}

where $\omega_D$ is the detuning of the coherent drive from the mmwave cavity. For a thermal state input of $n_{th}$ photons
\begin{equation}
    \overline{g}_{th}^{(1)}(\tau) = \frac{n_{th}}{2\pi} \int_{-\infty}^{\infty}\left.|S(\omega)|^2\right|_{\kappa_{mm}^{ext}=\kappa_{mm}} e^{-i \omega \tau} d\omega
\end{equation}
For a coherently displaced thermal state, the rate at the output is just the sum of the individual coherent and thermal contributions. For the theory curves in Fig.~\ref{fig:coherences}, we calculate $\overline{g}^{(1)}(0)$ and scale it by the experimentally measured loss in the optical path. 

It can be further shown that for a displaced thermal state, the second order correlation function, \mbox{$g^{(2)}(\tau)\equiv\frac{\langle a^\dagger(t)a^\dagger(t+\tau)a(t+\tau)a(t)\rangle} {\langle a^\dagger(t)a(t)\rangle\langle a^\dagger(t+\tau)a(t+\tau)\rangle}$}, is given by :

\begin{equation}
\label{eqn:full g2}
    g^{(2)}(\tau) = 1+\frac{|\overline{g}_{th}^{(1)}(\tau)+\overline{g}_{coh}^{(1)}(\tau)|^2-|\overline{g}_{coh}^{(1)}(\tau)|^2}{|\overline{g}_{th}^{(1)}(0)+\overline{g}_{coh}^{(1)}(0)|^2}
\end{equation}

Note that in the absence of a coherent drive this reduces to the expected $1+|g^{(1)}(\tau)|^2$, where $g^{(1)}(\tau)$ is now the \emph{normalized} first order correlation function.

\subsection{Projected conversion efficiency with improvements}
\label{SI:improvements}

The main limitation to conversion efficiency is the blue cooperativity, $C_b$, since the other cooperativities can be increased either by increasing the atom number or increasing the Rydberg state dressing fraction by reducing the detuning $\Delta$. Some simple changes can significantly increase the conversion efficiency. As stated earlier in Methods~\ref{SI:optical pumping}, the choice of Rydberg states and mmwave polarization, while first made for simplicity, is not ideal for transduction. Having a mmwave polarization orthogonal to the optical cavity axis would allow circular polarizations for all the fields while maintaining phase matching. This would increase the blue Rabi frequency, $\Omega_b$ by a factor of 2. Raman sideband colling of the atoms would allow us to significantly decrease the coherence decay rate of the collective $36\textrm{S}_{1/2}$ state, $\Gamma_R$, so that it is limited only by the decay of the Rydberg state. Cooling the atoms would also allow the cloud to be compressed by loading into an intra-cavity optical lattice or dipole trap, allowing us to significantly decrease the blue beam size. We estimate that these improvements would allow the blue cooperativity to be increased to about 2400 from the current 30, leading to an efficiency of $1-2/\sqrt{C_b} \approx 0.96$. We also note further increases are possible if a quasi-phase matching scheme is adapted (see SI~\ref{SI:qpm}), which would allow implementation of a running wave blue optical cavity.

\subsection{Overall conversion efficiency}
The overall conversion efficiency is given by Eqn.~\ref{eqn:efficiency}. For us, this is primarily limited by low external coupling rate of the mmwaves, $\kappa_{mm}^{ext} \approx 2\pi\times 55$~kHz, which is more than a factor of 10 smaller than the overall $\kappa_{mm}$, set primarily by the internal losses of the mmwave cavity. This limits our overall conversion efficiency to $\approx 2.5\%$. This $\kappa_{mm}^{ext}$ is primarily a function of the length and the diameter of the incoupling waveguide, but in our case the final value was obtained by adding a thin sheet of copper with a small hole in it to reduce it even further (by a factor of 2). We chose this to minimize $\kappa_{mm}$ and enable experiments that explore the single particle mmwave non-linearity, in addition to experiments with quantum transduction. In fact, $\kappa_{mm}^{ext}$ can be easily increased to $100$-fold its current value by removing the copper sheet and shortening the length of the incoupling waveguide by $2$~mm. The corresponding decrease in the mmwave cooperativity $C_{mm}$ can be compensated by increasing the $35\textrm{P}_{1/2}$ state dressing fraction. Although a more desirable regime for operation of the transducer would be lowering the temperature to $1$ K, which would reduce the internal loss rate of the cavity to $< 2\pi\times10$~kHz~\cite{suleymanzade2020tunable}. At this point the external coupling can be increased by a factor of 10, resulting in roughly the same $\kappa_{mm}$ and conversion bandwidth as we have now.

\begin{figure*}
\centering
  \includegraphics[width=1\linewidth]{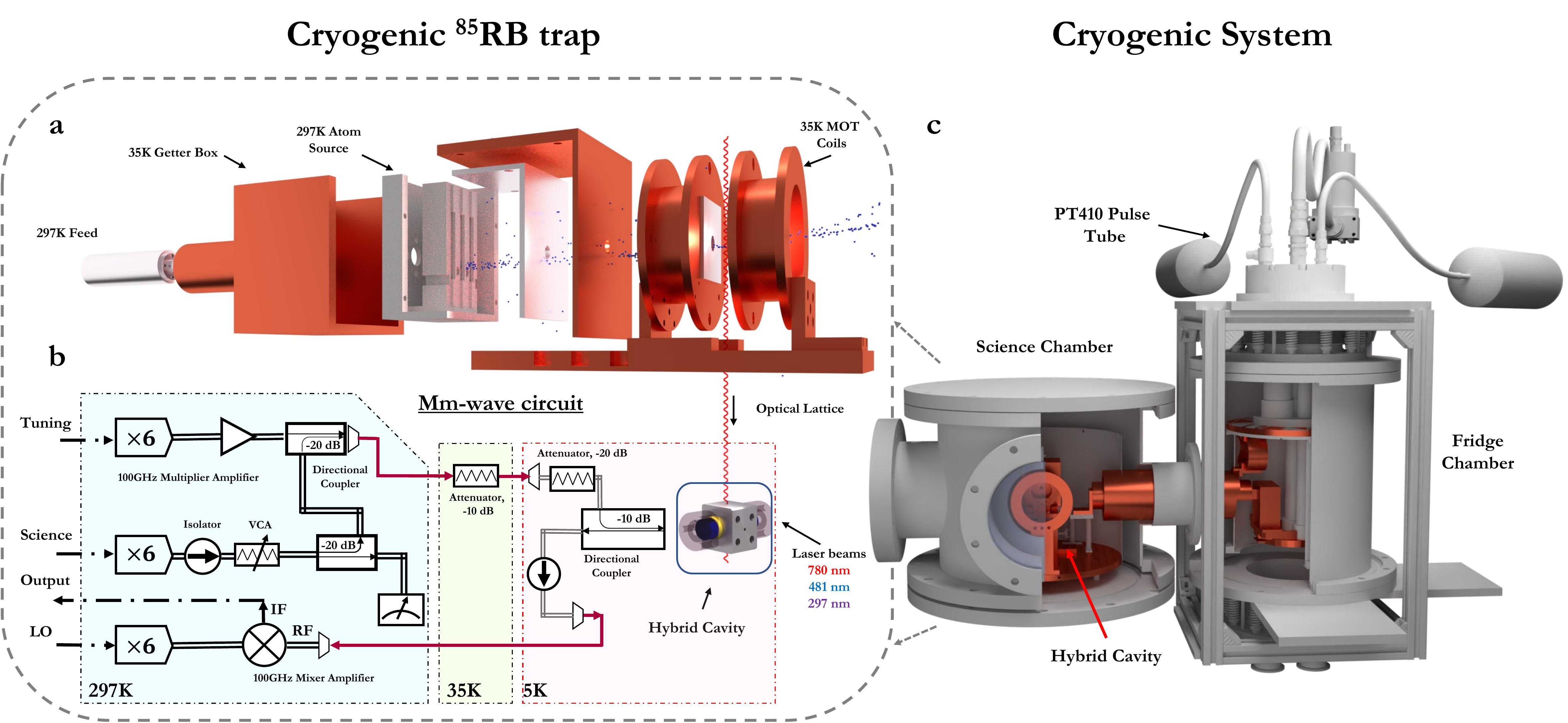}

\caption{\textbf{Cryogenic System and mmwave control.} \textbf{a,} Cryogenic $^{85}$Rb trap. The atoms are emitted from a source mounted at room temperature, then propagate through a small aperture at the back of the grating towards the capture volume of the grating MOT. The Helmholtz coils and all the supporting structures are thermalized to 40K.
\textbf{b,} Millimeter wave circuit used for characterizing the superconducting cavity and sending in mmwave photons for the inter-conversion experiment. Most of the power from the ``science" source is diverted towards a power detector, which is used to actively stabilize the mmwave power using a voltage controlled attenuator (VCA). Both the science and the tuning sources are equipped with rf-switches to completely turn off any signal to the cavity.
\textbf{c,} The custom two-chambered 4K cryogenic system built for our hybrid cavity-QED experiments with Rydberg atoms.
    }
\label{fig:SupFig1}
\end{figure*}

\begin{figure*}
\centering
  \includegraphics[width=1\linewidth]{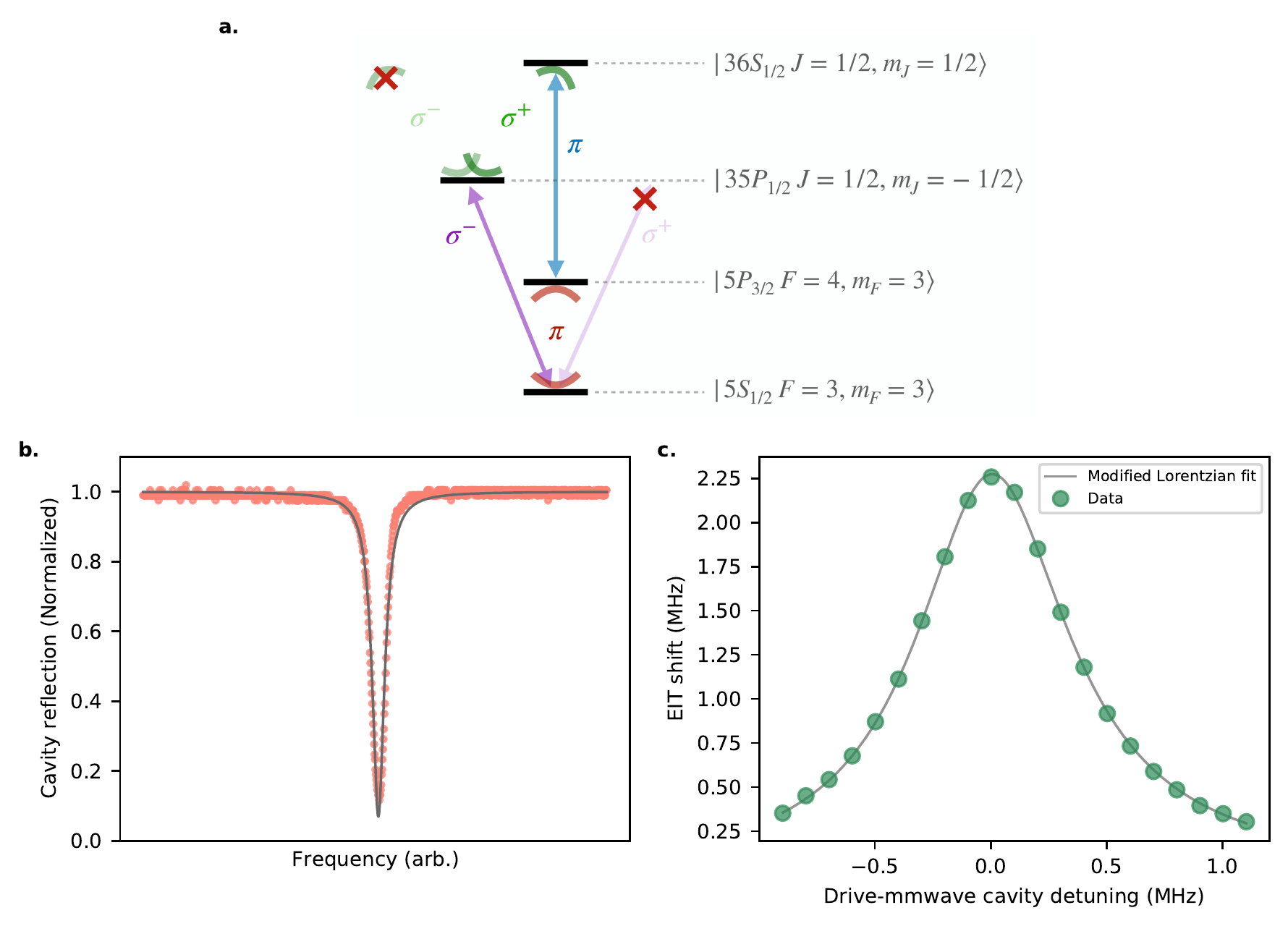}
\caption{
\textbf{a,} The exact levels and light polarizations involved in transduction. The mmwave cavity (green) science mode is linearly polarized along the optical cavity (red) direction. We choose (through a magnetic field and optical pumping) the quantization axis along the lattice beam direction to achieve phase matching and zero angular momentum change on a round trip at the same time. The mmwave mode polarization, which is orthogonal to this quantization axis can be decomposed as a linear combination of $\sigma^+$ and $\sigma^-$ polarizations. The UV beam (purple) is linearly polarized orthogonal to both the lattice direction and the optical cavity axis, and thus can also be decomposed as a linear combination of $\sigma^+$ and $\sigma^-$ polarizations. The blue beam and the $780$~nm probe (or the emitted interconverted photon) are linearly polarized along the lattice direction, and therefore have $\pi$ polarizations. We start by optically pumping to the stretched hyperfine state, which results in the UV $\sigma^+$ component and the mmwave $\sigma^-$ component not coupling to any available state. The effect of the Raman couplings generated from these polarizations is suppressed due to the magnetic field lifting the degeneracy.
\textbf{b,} Reflection measurement of the port of the optical cavity at which the photons are counted for any of our experiments described in the main text. The minimum reflection point directly gives $(1-2\kappa_{opt}^{ext}/\kappa_{opt})^2$. The gray line is the theory fit, with slight deviations due to a small polarization splitting of the cavity, and polarization impurity of the probe. 
\textbf{c,} ``In-situ" measurement of the mmwave cavity spectrum using the shift of the dark polariton/EIT resonance. The atomic transition is far-detuned ($2\pi\times12.4$~MHz) from the cavity and a mmwave drive is scanned in frequency around the \emph{cavity} resonance. The resulting ac-Stark shift of the dark polariton is porportional to the mmwave power in cavity. The gray line is a fit to a Lorentzian modified to account for the effect of the changing detuning of the drive from the atoms.
    }
\label{fig:sup levels}
\end{figure*}

\begin{figure*}
\centering
  \includegraphics[width=1\linewidth]{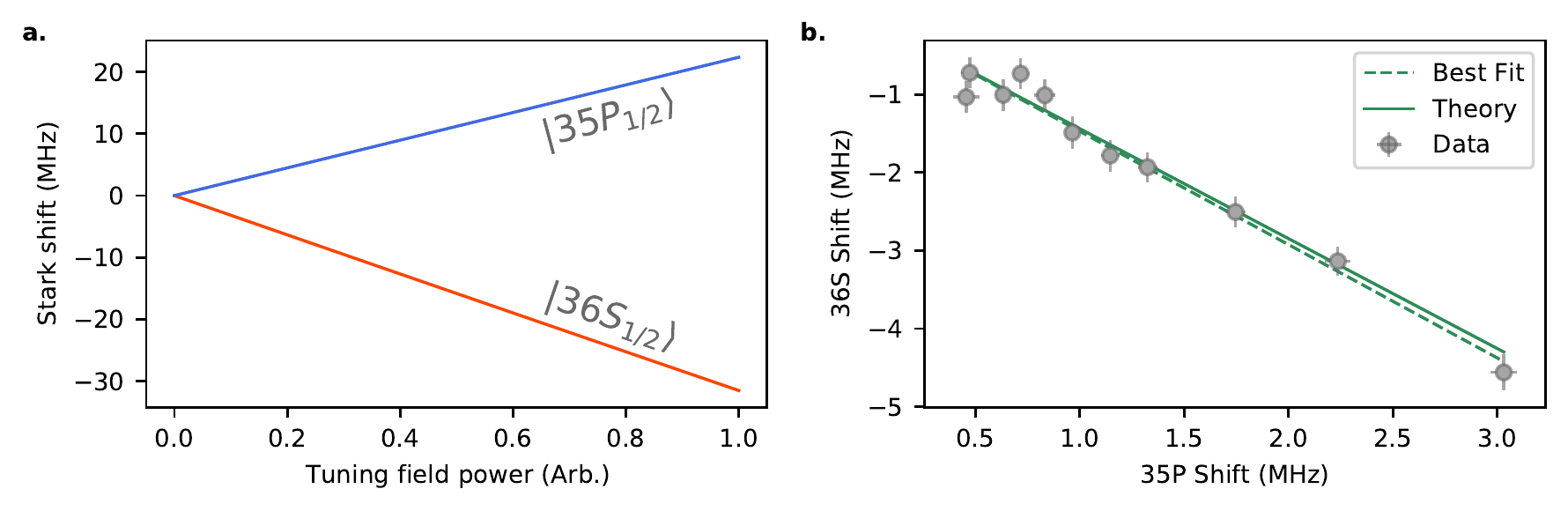}
\caption{
    \textbf{Stark tuning of the atomic states}. \textbf{a,} The calculated ac Stark shifts of the $36\textrm{S}_{1/2}$ and $35\textrm{P}_{1/2}$ ``science" states as the power in the $101.318$~GHz ``tuning" mode of the cavity is increased. The tuning of the atomic states allows us to control the detuning ($\Delta$) between the atomic transition and the mmwave cavity for transduction, as well as other experiments like those in Fig.~\ref{fig:syscharacterize}c. \textbf{b,} As we vary the classical drive on tuning mode, we measure the shifts of the $36\textrm{S}_{1/2}$ and $35\textrm{P}_{1/2}$ states using a cavity Rydberg EIT and direct UV spectroscopy from the $5\textrm{S}_{1/2}$ state, respectively. This calibration then allows us to infer the $35\textrm{P}_{1/2}$ shift (which is more involved to measure day-to-day) from the $36\textrm{S}_{1/2}$ shift measured via cavity Rydberg EIT. We find excellent agreement between our calculation and the observed shifts.
	}
\label{fig:sup tuning}
\end{figure*}

\begin{figure*}
\centering
  \includegraphics[width=1\linewidth]{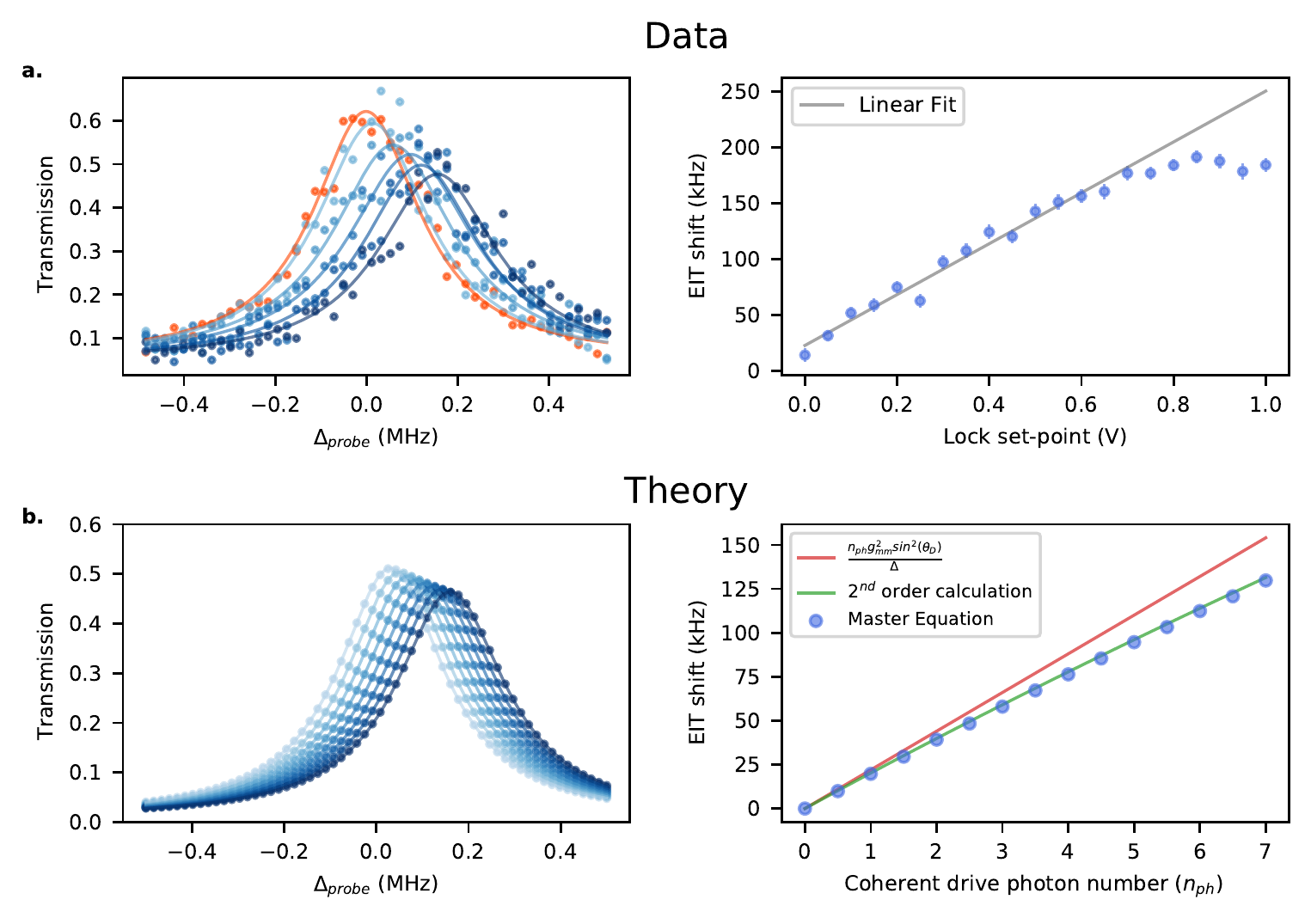}
\caption{\textbf{Measuring intracavity photon number ($n_{ph})$ from dispersive shifts}. \textbf{a,} Left - Measured dark polariton/Rydberg EIT spectra as the ``science" mode drive strength is increased at an atom-mmwave cavity detuning, $\Delta=2\pi\times1.4$~MHz. We first measure a reference spectrum with the mmwave source entirely turned off using an rf-switch (orange). An actively locked mmwave drive (using a power detector and a VCA, see Methods~\ref{SI:apparatus}, Fig.~\ref{fig:SupFig1}) is then applied with increasing strength (light blue to dark blue). The solid lines are Lorentzian fits. Right - The extracted shifts by fitting the spectra in the left panel as the lock set point is increased. Note that even at the lowest lock set-point, there is some leakage from the VCA, which necessitates measuring the orange reference in the left panel with no mmwave power from the source. After a threshold, the shifts saturate because there is the VCA reaches lowest attenuation and there is no more mmwave drive power available. 
\textbf{b,} Master equation calculation of the data in $\textbf{a}$ using the non-linear Hamiltonian given by equation~\ref{eqn:non lin ham} with parameters independently measured or calculated from first principles ($g_{mm}$ and number of thermal photons) and an added coherent drive. This calculation allows us to calibrate our coherently driven photon number ($n_{ph}$) and the dispersive shift of the EIT resonance. Left - The calculated spectra with Lorentzian fits. Right - The shift of the EIT resonance against the number of mmwave photons driven into the cavity by the coherent drive in steady state. The simple first order expression (red) over-estimates the shifts. We instead calculate and use a much more accurate analytical result (green) in SI~\ref{SI:dispersive shift}.
}
\label{fig:sup dispersive}
\end{figure*}

\renewcommand\arraystretch{1.5}

\begin{table*}[htb]
\centering
\begin{tabular}{ |l c l| } 
\hline

\textbf{Parameter} & \textbf{Symbol} & \textbf{Value} \\ 
\hline
Millimeter wave cavity ``science" mode frequency & -- & $2\pi\times99.42376(1)$~GHz \\ 
\hline
Millimeter wave cavity ``tuning" mode frequency & -- & $2\pi\times101.318$~GHz  \\ 
\hline
$35\textrm{P}_{1/2} \leftrightarrow36\textrm{S}_{1/2}$ transition frequency without stark tuning & -- & $2\pi\times99.436$~GHz  \\
\hline
Single atom-optical cavity coupling & $g_{opt}$ & $2\pi\times206$~kHz  \\ 
\hline
Optical cavity linewidth & $\kappa_{o}$ & $2\pi\times1.71(4)$~MHz  \\ 
\hline
Out-coupling rate of the measurement optical cavity port & $\kappa_{o}^{ext}$ & $2\pi\times1.07(3)$~MHz \\ 
\hline
Decay rate of the $5\textrm{P}_{3/2}$ state & $\Gamma$ & $2\pi\times6.065$~MHz\\
\hline
Blue beam Rabi frequency & $\Omega_b$ & $2\pi\times1.45(5)$~MHz\\
\hline
Decoherence rate of the $36\textrm{S}_{1/2}$ collective state & $\Gamma_R$ & $2\pi\times56(8)$~kHz\\
\hline
Single atom-mmwave cavity coupling & $g_{mm}$ & $2\pi\times182$~kHz\\
\hline
Millimeter wave cavity linewidth & $\kappa_{mm}$ & $2\pi\times805(5)$~kHz\\
\hline
Millimeter wave cavity external coupling rate & $\kappa_{mm}^{ext}$ & $2\pi\times55(5)$~kHz\\
\hline
UV beam effective Rabi frequency& $\Omega_{UV}$ & $2\pi\times230(3)$~kHz\\
\hline
Optical path efficiency including SPCM efficiency & $f_o$ & 0.28(2)\\
\hline

\end{tabular}

\caption{Table of key experimental parameters}
\label{table:parameter table}
\end{table*}

\renewcommand{\tocname}{Supplementary Information}
\renewcommand{\appendixname}{Supplement}

\setcounter{equation}{0}
\setcounter{figure}{0}
\renewcommand{\theequation}{S\arabic{equation}}
\renewcommand{\thefigure}{S\arabic{figure}}

\incltocpage
\clearpage
\tableofcontents
\appendix
\section{Experimental Apparatus}
\label{SI:apparatus}
\subsection{Cryogenic System for cold atoms}
 The home-built cryogenic vacuum system is designed to provide ample optical access and vibration isolation for the hybrid cavity-QED experiment at 4K. It consists of the fridge and science chambers, shown in Fig.~\ref{fig:SupFig1}\textbf{c}, which are connected horizontally to minimize the propagation of vertical vibrations from the dry pulse tube into the experimental chamber. The fridge chamber contains the \textit{Cryomech PT410 CPA289C series} dry pulse tube refrigerator, turbopump for initial pumpdown, and vibration isolation system. The room temperature component of the cryocooler is supported independently by a frame hanging from above, and is connected via bellows to the outer vacuum can of the fridge chamber, which rests on the optical table. The science chamber contains mmwave circuitry, atom sources, viewports, and optics needed to perform the experiments. The two chambers are rigidly connected at room temperature through mechanical flanges of the chamber walls and bolted to the optical table. The internal subsystems at cryogenic temperatures (35 K and 4 K) are attached through a cylindrical cold arm with flexible copper braids and foil connectors. We thermalize everything we can to the 35 K stage, since it has up to 30 W of cooling power, and only thermalize our superconducting hybrid cavity with the last stage of the mmwave circuit to the 4 K stage, which provides a few $100$s of mW in the fully loaded system.
 
 The cryogenic Magneto-Optical Trap, mounted on the 35 K plate, consists of two copper quadrupole coils, the thermally isolated box with $^{85}$Rb dispensers and grating chip. A grating MOT design was chosen to minimize the required number of beams and windows to trap large enough cloud of atoms. The chip is a $20\,\textrm{mm} \times 20$~mm lithographically patterned Al grating on Si wafer with a pitch $d = 1400$~nm, which corresponds to the first-order reflection at $34$ degrees at $780$~nm. It was designed and built by the group in Strathclyde~\cite{mcgilligan2015phase, mcgilligan2017grating, nshii2013surface} specifically to trap Rubidium, by reflecting a single input beam into the tetrahedral pattern and creating a trapping volume at the centroid of the tetrahedron, $\approx 5$ mm off the surface of the grating. In order to avoid excessive heating and loss, the atomic source box is placed behind the grating, and atoms enter the trap volume through a laser-cut aperture of 5 mm diameter, as shown in Fig.~\ref{fig:SupFig1}\textbf{a}.
 
To optimize the efficiency of the experimental sequence the atomic \textit{SAES getters} are run continuously at $\approx3.5$~A throughout the operation of the machine. We can measure the effective atom number in the optical cavity mode and from that extrapolate the atom numbers in the lattice and the MOT through fluorescence imaging of the atomic cloud. Our typical atom numbers and temperatures are detailed in the following table:

\begin{tabularx}{0.4\textwidth} { 
  | >{\raggedright\arraybackslash}XXX
  | >{\centering\arraybackslash}XXX
  | >{\centering\arraybackslash}XXX | }
 \hline
 \textbf{Exp step} & \textbf{\# of atoms} & \textbf{Temp ($\mu$K)} \\
 \hline
 MOT  & $\approx 50\times10^3$  & $\approx 100 $  \\
 PGC  & $\approx 50\times10^3$  & $\approx 4 $  \\
 Lattice  & $\approx 3500$  & $\approx 4 $  \\
 Cavity  & $\approx 50-1500$  & $\approx 4 $   \\
\hline
\end{tabularx}

\subsection{Millimeter-wave resonator}
\label{SI:mmwave resonator}

\begin{figure*}
\centering
  \includegraphics[trim = 0cm 16cm 0cm 0cm]{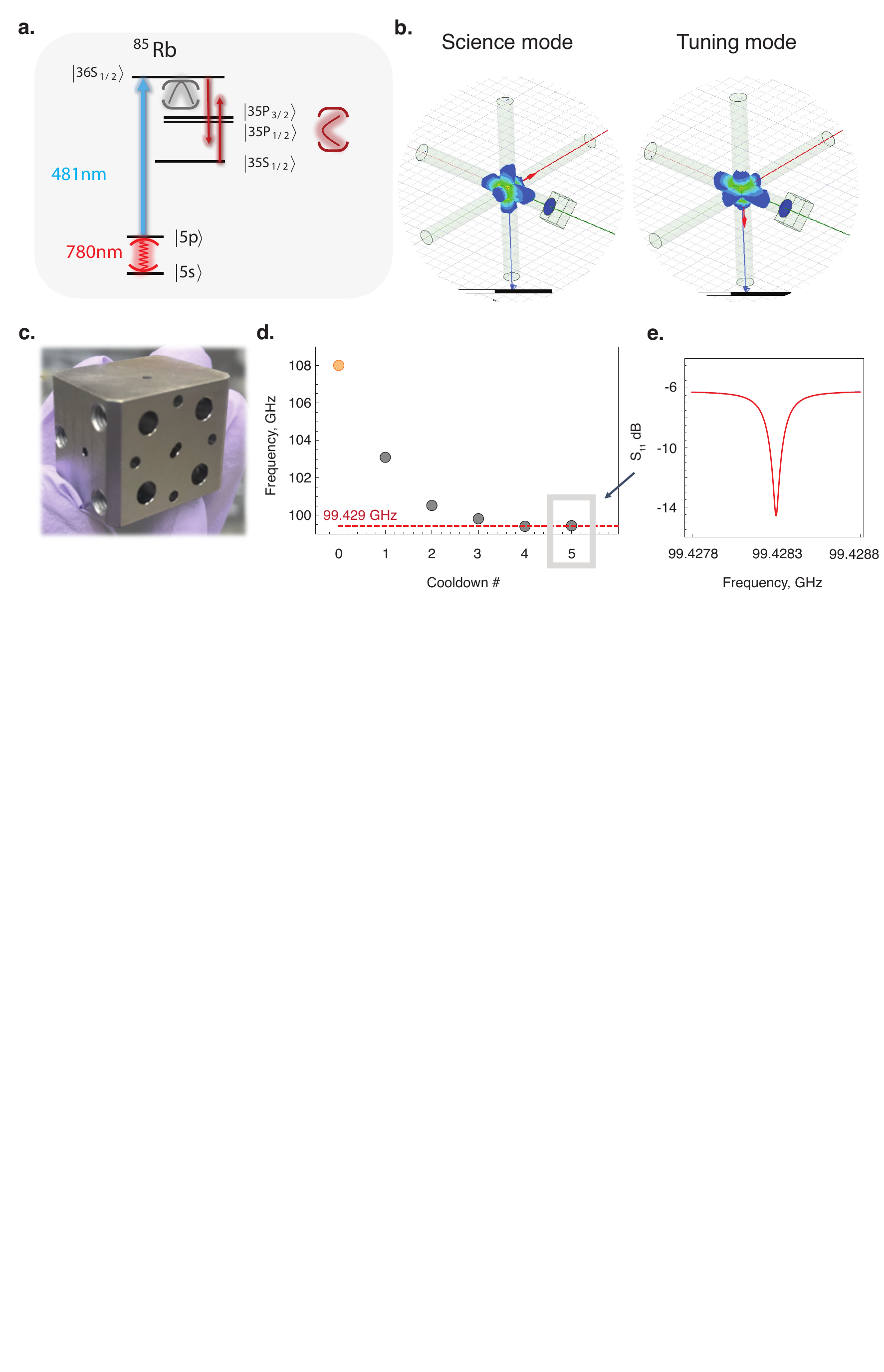}
\caption{Mmwave resonator modes: \textbf{a.} The energy levels involved in tuning the atomic resonances to the lowest science mode of the mmwave cavity, shown in gray, using the next lowest tuning mode, shown in dark red. \textbf{b.} HFSS simulation of E field geometry of the science and tuning modes in the resonator. \textbf{c.} The picture of the machined Nb mmwave cavity before attaching the optical cavity mirrors. \textbf{d.} Frequency of the science mode after rounds of etching and squeezing. After the last round of squeezing, the frequency of the mmwave cavity was within $1$~MHz of the bare Rydberg transition of $99.429$~GHz, but it eventually settled to  $99.42376$~GHz when mounted in the main experimental chamber. Note: In the experiment we used Zeeman shift to set the atomic transition frequency to be $99.436$~GHz. \textbf{e.} The reflection spectrum of the lowest mode of the mmwave cavity after the tuning process at $1$~K. The figure is incorporated from~\cite{suleymanzade2021millimeter}.    
    \label{fig:Mmcavmodes}}
\end{figure*}

The mmwave resonator~\cite{suleymanzade2020tunable} is composed of three orthogonal intersecting tubes that act as evanescent waveguides at the ``science" and ``tuning" frequencies of our hybrid system, as shown in Fig.~\ref{fig:Mmcavmodes}\textbf{a,b,c}. The tube diameters are chosen such that the lowest mode of the cavity is close to the (Zeeman shifted) transition between Rydberg states $\ket{36 \textrm{S}_{1/2}}$ and $\ket{35 \textrm{P}_{1/2}}$ at $\approx 99.43614$~GHz. A higher tuning frequency of $101.318$~GHz is chosen for ac Stark shifting the respective atomic levels into resonance with the science mode, as shown in Fig.~\ref{fig:Mmcavmodes}\textbf{a} and Fig.~\ref{fig:sup tuning}\textbf{a}. The E field orientations of the two lowest cavity modes are orthogonal at the location of atoms -- along the optical cavity direction for the science mode and along the optical lattice direction for the tuning mode. The lengths of tubes are chosen to suppress the leakage of light, except through the coupling port, which is intentionally made shorter to set the coupling quality factor to be in the range: $Q_{couple} = 10^5 - 10^6$. 

\begin{figure*}
\centering
  \includegraphics[trim = 0cm 15cm 0cm 0cm]{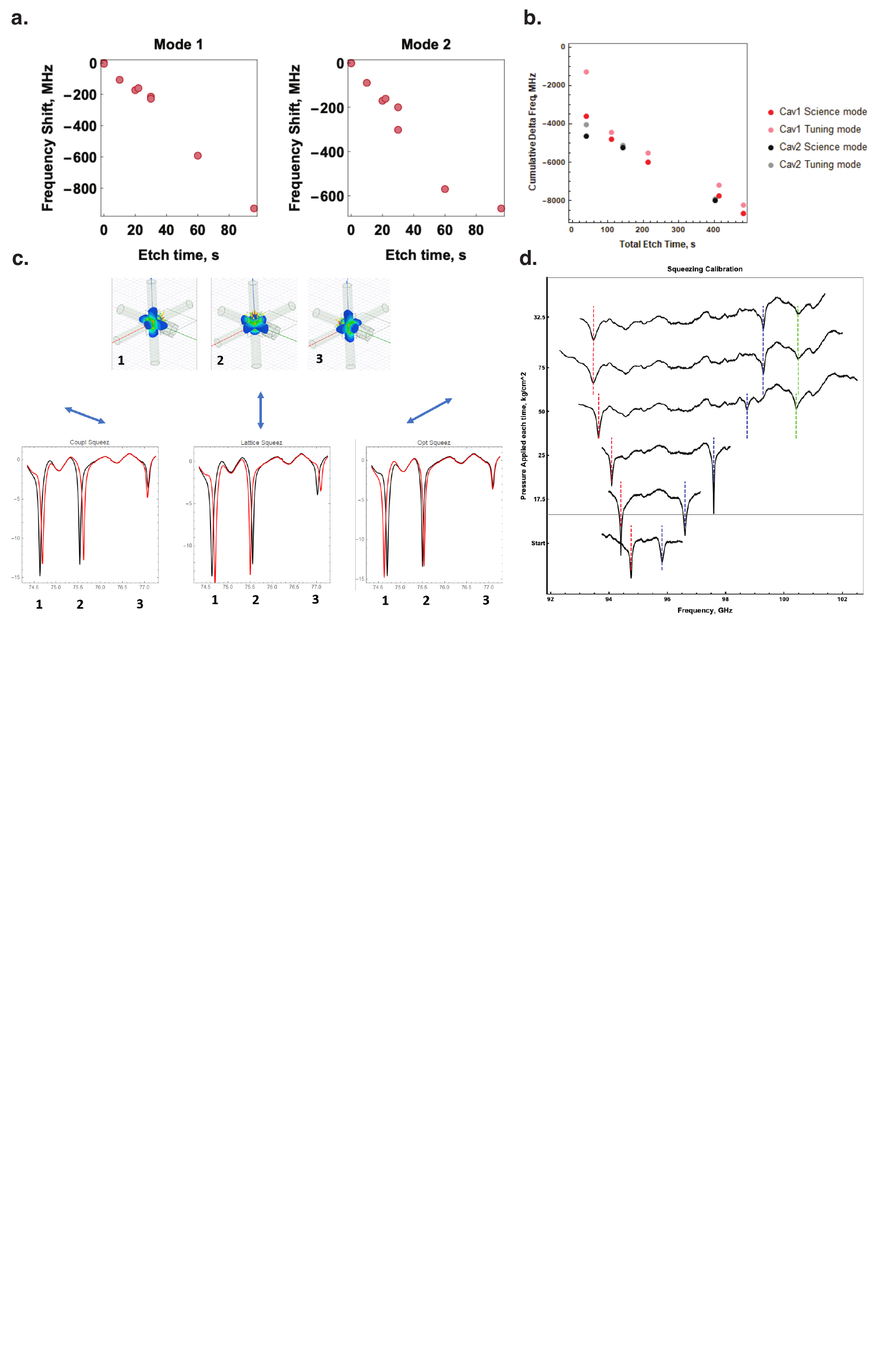}
\caption{Millimeter wave resonator tuning mechanisms: \textbf{a.} Calibration data for the etching rate of the mmwave cavity. \textbf{b.} The cumulative change in frequency of the modes on two different mmwave resonators after total etch time. The early etch times tend to be more violent due to sharp corners left from machining. \textbf{c.} Mechanism of mechanical squeezing of the mmwave cavity along three cavity directions. \textbf{d.} Calibration data for the squeezing rate of the mmwave cavity. The figure is incorporated from~\cite{suleymanzade2021millimeter}
    \label{fig:MmcavTuning}}
\end{figure*}

The cavity is manually machined out of a block of high purity cavity-grade Nb using carbide drills followed by reamers for a smoother finish. The diameters of drills are chosen to be significantly undersized. We also leave a few extra GHz ($\approx 0.003$") extra material for acid etch cleaning and tuning. The machined mmwave cavity is shown in Fig.~\ref{fig:Mmcavmodes}\textbf{c} with hole patterns for optical mirror assembly on the sides and the WR10 waveguide flange pattern for the mmwave port. Next, the cavity undergoes an acid etch in a bath of 2:1:1 solution of $H_3 PO_4$: $H NO_3$ : $HF$, with the etching rate calibrated on test cavities, as shown in Fig.~\ref{fig:MmcavTuning}\textbf{a, b, c}. After initial rounds of coarse etching with a rate of 6$\mu$m/min, the cavity is mechanically squeezed using a hydraulic press in addition to finer acid etching steps to achieve the required frequency precision as shown in Fig.~\ref{fig:Mmcavmodes}\textbf{d}. The tuning vs squeezing force is also calibrated on test devices, as shown in Fig.~\ref{fig:MmcavTuning}\textbf{c}.

The physical dimension of the mmwave cavity that controls the resonant frequency of $\omega\approx 2\pi\times 100$~GHz is $L\approx 1.4$~mm; to achieve efficient transduction, we need this resonator to be within the collectively enhanced $g_{mm}$, of a few MHz, corresponding to a size accuracy requirement on the resonator dimensions of $\delta L=L\times \frac{1\textrm{MHz}}{100\textrm{GHz}}\approx 14$~nm. The repeated process of etch-tuning, squeezing, and measuring the resonance at 4K described above allows us to achieve this accuracy, which is far beyond conventional machining techniques ($\sim 1/1000$~inch $\approx 25\mu$m). The frequency of the science mode eventually settled to $99.42376$~GHz in the final cooldown, where experiments were performed. Subsequent control over the atom-cavity detuning is achieve by ac Stark shifting the atomic levels by driving the tuning mode. This is detailed in Methods~\ref{SI:tuning}.

\subsection{Vibration-insensitive cryogenic optical cavity}
\label{SI:optical cavity}

\begin{figure}
\centering
  \includegraphics[width=0.5\textwidth]{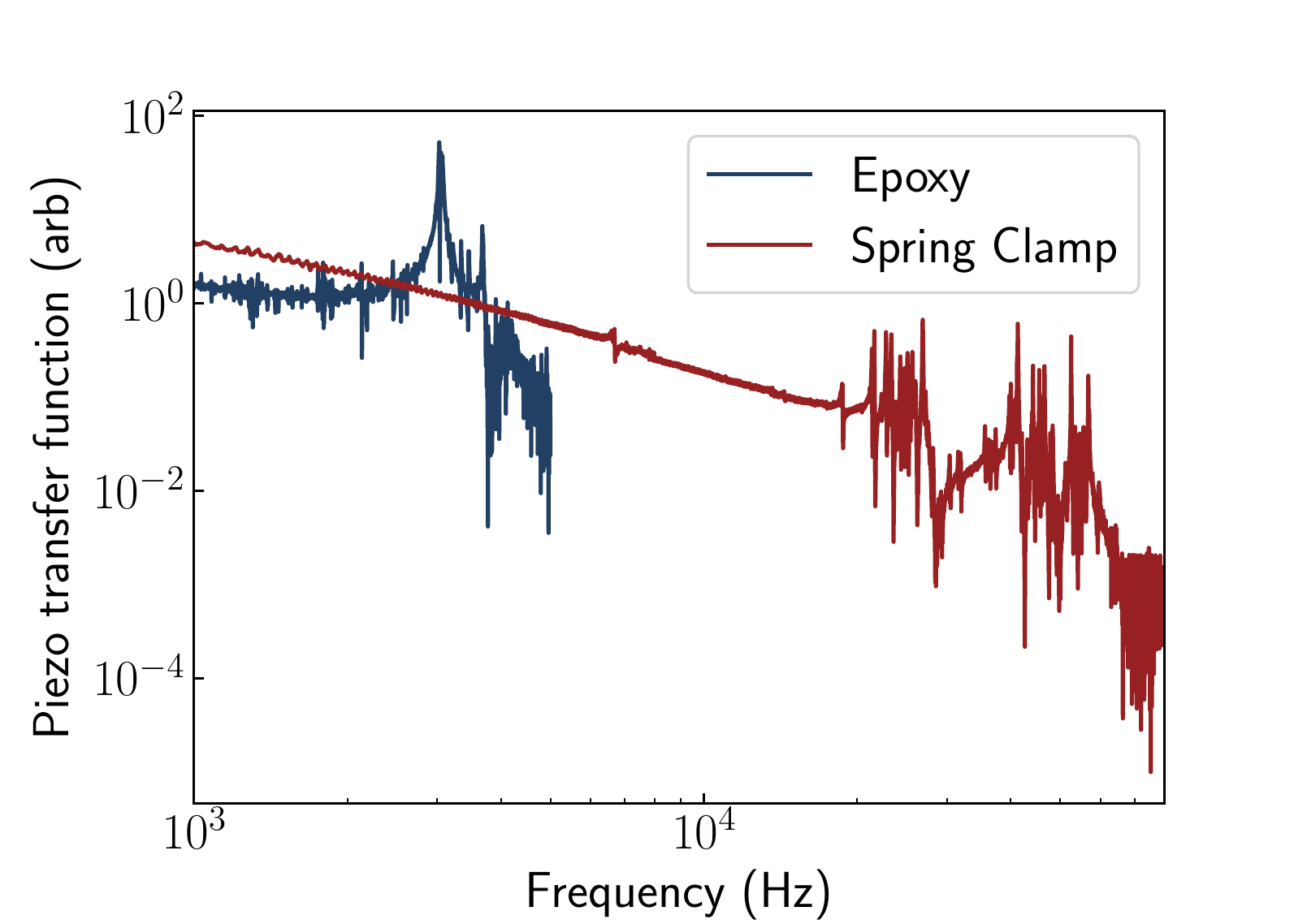}
\caption{Mechanical transfer function for optical cavities with epoxy and spring clamp construction.
    \label{fig:transferfunction}}
\end{figure}
The optical cavity is a two mirror Fabry-P\'erot resonator designed for compatibility with a closed-cycle cryocooler and a 3D mmwave cavity. The mirrors (7.75 mm diameter, 4 mm thickness, 25 mm radius of curvature) are attached to Thorlabs PA44LEW piezoelectric actuators, which are themselves attached directly to the mmwave cavity, so that the optical mode passes through holes in the niobium. This makes an optical cavity with a length of about $2.54$~cm and a mode waist of approximately $55$~microns. The reflectivities measured by the manufacturer (Layertec) were $0.9994$ and $0.9988$ for the two mirrors. We use the lower reflectivity mirror as the output port of the cavity for transduction and other measurements. These reflectivities are consistent with our measured finesse ($\approx 3450$) and the measured $\kappa_{opt}^{ext}/\kappa_{opt}\approx0.63$ (Fig.~\ref{fig:sup levels}b).

Vibrations from closed-cycle cryocoolers may couple mechanically to an optical cavity, complicating the task of maintaining length stability for resonance ($\Delta L<\lambda/(2\mathcal{F})$, where $\mathcal{F}$ is the finesse). Although cavities assembled with epoxy are sufficiently rigid given moderate $\mathcal{F}\sim3000$ and a well-design experimental chamber, we found that the epoxy bond would loosen and develop large mechanical resonances over one to several cooldown cycles, greatly increasing the impact of vibrations. The best cryogenic epoxy we tested is MasterBond EP21TCHT-1, but even this loosens unpredictably, leading to resonances. Creating a vibration-insensitive cavity amounts to maximizing the frequency of-, and minimizing the response at-, all mechanical resonances.

We replace epoxy with a clamp and spring to press the mirror, piezo, and niobium block together, as shown in Fig.~\ref{fig:setup}. A modest spring constant is soft with respect to extensions of the piezo and thermal contraction of the clamp, but stiff for relative motion between the mirror and cavity spacer, where the spring stiffness is added to the compressional stiffness of the piezo and niobium.

The design consists of the piezo and mirror, as well as a guidepin (nitronic 60 steel), two stacked conical disc springs (individually rated for 0.13 mm deflection and 245 N load), and an aluminum housing tube. One end of the guidepin is flared and has a depression to fit the back face of mirror. The other end is a long, hollow tube, which passes through the two disc springs (stacked in the same orientation) and a central hole in the housing tube. This serves to center the mirror and disc springs relative to the housing. The inside length of the housing equals the height of the stack (measured precisely with a height gauge), minus the desired deflection of the springs, so that the back face clamps down the springs when the housing is screwed into the niobium block. All components have central holes to allow light through the structure.

The vibrational performance of the spring clamp is consistent across many cooldowns. Its mechanical transfer function is compared against cryogenic epoxy in Fig.~\ref{fig:transferfunction}. Compared to the large, single-pole mass-on-a-spring resonance from the mirror mass and weakened epoxy stiffness, the spring clamp shows several pole-zero pairs, as expected from Foster's reactance theorem. We believe these resonances arise from self-resonances of the springs or, more likely, the housing tube. As they are smaller and higher frequency than the epoxy resonance, they couple in far less vibrational noise from the environment.

At cryogenic temperatures, the rms cavity length excursion is $<25$ pm, sufficient for a moderate finesse cavity. A shorter mirror (and therefore housing tube) pushes the first major resonance from 20 kHz to 40 kHz, which will substantially improve performance, as would moving to a monolithic flexure spring mount.

Vibrations in the cavity decay much faster at room temperature than at 4 K for both epoxy and spring designs, and the 4 K transfer function shows higher quality factor resonances. This is likely due to the increased acoustic quality factor of most materials at low temperatures. In addition, the piezo throw decreases to about 15\% of its room temperature value, so we connect both piezos electrically in parallel to achieve a full free spectral range of travel.

\section{Theory}
\label{SI:interconversiontheory}
This section presents the full Hamiltonian for a sample of atoms coupled to both optical and mmwave cavities, and derives a simpler model using ``collective states", where the excitations of a large number of atoms can be treated as approximately bosonic. To build intuition, two different derivations are presented. The first approach begins with the full, time-dependent Hamiltonian, which is transformed into the frame of the dressed states (superpositions $5\textrm{S}_{1/2}$ and $35\textrm{P}_{1/2}$ states) created by the UV drive. Neglecting the off-resonant dressed state and introducing collective excitation operators from the reservoir established by these dressed states directly leads to a linearized interconversion Hamiltonian. In the second approach, we introduce collective excitations from the beginning, now from the reservoir of all the atoms initially in the $5\textrm{S}_{1/2}$ ground state. This enables us to explicitly derive a non-linear Hamiltonian without the UV drive, which is critical to quantitatively understanding some results of the experiment, like the Purcell broadening of the dark polariton when the atoms are tuned close to the cavity resonance, as well as the dispersive shift of the polariton used to measure the intra-cavity photon number. The small non-linearity can then be neglected in presence of the UV drive, leading to the same linear interconversion Hamiltonian as the first approach. 

For both these approaches we initially ignore the standing wave nature of the optical cavity, assume a running wave coupling and derive the main results concerning interconversion in following sections by solving the Heisenberg-Langevin equations. We then introduce the effect of the other non-phase matched running wave in Section~\ref{SI:standing wave}. Finally we show how, in the limit that dark and bright polaritons are spectrally well separated, the linear interconversion Hamiltonian can be reduced to a beam-splitter between the dark polariton and the mmwave mode.

\subsection{Deriving a simplified linear Hamiltonian}
The full Hamiltonian for the system in the rotating wave approximation is:
\begin{multline}
H=\omega_{a}\hat{a}^{\dagger}\hat{a}+\omega_{b}\hat{b}^{\dagger}\hat{b}+\sum_{i}\multiL\omega_{e}\hat{\sigma}_{ee}^{i}+\omega_{r}\hat{\sigma}_{rr}^{i}+\omega_{f}\hat{\sigma}_{ff}^{i}\\
+\biggl(g_{opt,i}\hat{a}\hat{\sigma}_{eg}^{i}+\Omega_{b,i}e^{-i\omega_{blue}t}\hat{\sigma}_{re}^{i}\\
+g_{mm,i}\hat{b}\hat{\sigma}_{rf}^{i}+\Omega_{uv,i}e^{-i\omega_{uv}t}\hat{\sigma}_{fg}^{i}+h.c.\biggr)\biggr]
\end{multline}

The subscripts $g$, $e$, $r$, and $f$ represent the $5\textrm{S}_{1/2}$, $5\textrm{P}_{3/2}$, $36\textrm{S}_{1/2}$, and $35\textrm{P}_{1/2}$ states respectively. The time dependence can be removed by applying a rotating-frame unitary transformation on the wavefunction, $\ket{\psi}\rightarrow e^{-iAt}\ket{\psi}$, where
\begin{multline*}
A=\omega_{o}a^{\dagger}a+\omega_{m}b^{\dagger}b\\+\sum_{i}\left[\omega_{o}\hat{\sigma}_{ee}^{i}+\left(\omega_{o}+\omega_{blue}\right)\hat{\sigma}_{rr}^{i}+\left(\omega_{o}+\omega_{blue}-\omega_{m}\right)\hat{\sigma}_{ff}^{i}\right]
\end{multline*}

Because there are two coupling paths between any two states in this Hamiltonian, time dependence can only be fully removed if the transformation satisfies $\omega_{o}+\omega_{blue}-\omega_{m}-\omega_{uv}=0$, which amounts to enforcing energy conservation around the transduction loop.

In addition, there is a propagation phase factor $e^{i\boldsymbol{k}\cdot\boldsymbol{r}}$ for each laser, which is implicit in the coupling constants $\Omega_{b,i}$, $\Omega_{uv,i}$, $g_{opt,i}$, $g_{mm,i}$ for an atom at position $\boldsymbol{r}_i$. These phase factors can be removed with an additional unitary transformation of the atomic wavefunctions, $\psi \rightarrow T \psi$, where $T=\sum_{i}\left[e^{i\boldsymbol{k}_{780}\cdot\boldsymbol{r}_i}\hat{\sigma}_{ee}^{i}+e^{i\left(\boldsymbol{k}_{780}+\boldsymbol{k}_{481}\right)\cdot\boldsymbol{r}_i}\hat{\sigma}_{rr}^{i}+e^{i\boldsymbol{k}_{uv}\cdot\boldsymbol{r}_i}\hat{\sigma}_{ff}^{i}\right]$.

Under these transformations, the Hamiltonian becomes

\begin{multline}
H=\delta_{a}\hat{a}^{\dagger}\hat{a}+\delta_{b}\hat{b}^{\dagger}\hat{b}+\sum_{i}\multiL\delta_{e}\hat{\sigma}_{ee}^{i}+\delta_{r}\hat{\sigma}_{rr}^{i}+\delta_{f}\hat{\sigma}_{ff}^{i}\\
+\biggl(g_{opt,i}\hat{a}\hat{\sigma}_{eg}^{i}+\Omega_{b,i}\hat{\sigma}_{re}^{i}+P_i g_{mm,i}\hat{b}\hat{\sigma}_{rf}^{i}+\Omega_{uv,i}\hat{\sigma}_{fg}^{i}+h.c.\biggr)\biggr]
\end{multline}
where $\delta_{a}=\omega_{a}-\omega_{o}$, $\delta_{b}=\omega_{b}-\omega_{m}$, $\delta_{e}=\omega_{e}-\omega_{o}$, $\delta_{r}=\omega_{r}-\omega_{blue}-\omega_{o}$, $\delta_{f}=\omega_{f}-\omega_{uv}$, and the phase-matching factor $P_{i}=e^{-i\left(\boldsymbol{k}_{481}+\boldsymbol{k}_{780}-\boldsymbol{k}_{mm}-\boldsymbol{k}_{uv}\right)\cdot\boldsymbol{r}_i}$. Phase matching occurs when $P_i=1$ for all atoms, for instance when all beams are copropagating, which will be assumed now.

Next, we move to a dressed-state picture by diagonalizing the strong UV coupling to the ground state atoms: $H_{uv}=\sum_{i}\left[\delta_{f}\hat{\sigma}_{ff}^{i}+\left(\Omega_{uv,i}\hat{\sigma}_{fg}^{i}+h.c.\right)\right]$. The dressed states have energies $E_{\pm}^i=\frac{1}{2}\left(\delta_{f}\pm\sqrt{\delta_{f}^{2}+4\left|\Omega_{uv,i}\right|^{2}}\right)$. So long as the splitting $E_+ - E_-$ is much larger than all other Rabi rates and linewidths, one dressed state can be far detuned and removed from all dynamics. This encompasses the case of large $\delta_f$ which corresponds to adiabatic elimination, but also the case of large UV Rabi rate. We thus drop the $\ket{+}$ state, leaving $\ket -_{i}  =\cos\frac{\theta_{i}}{2}\ket g-\sin\frac{\theta_{i}}{2}\ket f$, where $\tan\theta_{i}=\frac{2\Omega_{uv,i}}{\delta_{f}}$. We take all atoms in $\ket{-}$ as our vacuum state, and the Hamiltonian is
\begin{multline}
H=\delta_{a}\hat{a}^{\dagger}\hat{a}+\delta_{b}\hat{b}^{\dagger}\hat{b}+\sum_{i}\left[\delta_{e}^i\hat{\sigma}_{ee}^{i}+\delta_{r}^i\hat{\sigma}_{rr}^{i}+\left(\overline{g}_{opt,i}\hat{a}\hat{\sigma}_{e-}^{i}+h.c.\right)\right.\\
\left.+\left(\Omega_{b,i}\hat{\sigma}_{re}^{i}+h.c.\right)+\left(\overline{g}_{mm,i}\hat{b}\hat{\sigma}_{r-}^{i}+h.c.\right)\right] \label{eqn:dressedHamiltonian}
\end{multline}
where $\overline{g}_{opt,i}=g_{opt,i}\cos\frac{\theta_i}{2}$, $\overline{g}_{mm,i}=g_{mm,i}\sin\frac{\theta_i}{2}$, and we have gone into a frame where the atomic levels rotate at $E_i$. In this frame the detunings may differ between atoms; from here on we will assume uniform detunings, as for uniform $\Omega_{uv,i}$ or large $\delta_f$. We will also assume that $\Omega_{b,i}=\Omega_b$ is uniform for all atoms. These requirements will avoid coupling out of the collective state manifold.

We define collective excitation operators for spatial mode $l$ as $E^\dagger(l)=\frac{1}{l_c}\sum_i l_i\hat{\sigma}^i_{e-}$, $R^\dagger(l)=\frac{1}{l_c}\sum_i l_i\hat{\sigma}^i_{r-}$, where $l_c=\sqrt{\sum_i \left|l_i\right|^2}$. We also define $\Sigma_{re}=\sum_i \hat{\sigma}^i_{re}$, $N_e=\sum_i\hat{\sigma}_{ee}^{i}$, $N_r=\sum_i\hat{\sigma}_{rr}^{i}$. Then the Hamiltonian is
\begin{multline}
    H=\delta_{a}\hat{a}^{\dagger}\hat{a}+\delta_{b}\hat{b}^{\dagger}\hat{b}+\delta_{e} N_e +\delta_{r} N_r\\
    +\left(g_{opt,c} \hat{a}E^\dagger(g_{opt})+\Omega_b\Sigma_{re} + g_{mm,c} \hat{b} R^\dagger(g_{mm}) + h.c.\right)
\end{multline}
In the limit of low excitation number, these operators approximately obey bosonic commutation relations:
\begin{align*}
    \left[E(l),E^\dagger(m)\right]&=\left[R(l),R^\dagger(m)\right]=\left\langle l,m\right\rangle \\
    \left[E(l),E(m)\right]&=\left[R(l),R(m)\right]=0 \\
    \left[E(l),R^\dagger(m)\right]&=\left[E(l),R(m)\right]=0\\
    \left[\Sigma_{re}, E^\dagger(l)\right]&=R^\dagger(l) \\
    \left[\Sigma_{er}, R^\dagger(l)\right]&=E^\dagger(l) \\
    \left[\Sigma_{re}, E(l)\right]&=\left[\Sigma_{er}, R(l)\right]=0 \\
    \left[N_e, E(l)\right]&=E(l) \\
    \left[N_r, R(l)\right]&=R(l)\\
    \left[N_e, R(l)\right]&=\left[N_r, E(l)\right]=0
\end{align*}
where $\left\langle l,m\right\rangle=\frac{1}{l_c m_c} \sum_i l_i m_i$ represents the mode-matching between modes $l$ and $m$. It is now simple to obtain and solve the Heisenberg equations of motion for $\hat{a},\hat{b},E(g_{opt}),E(g_{mm}),R(g_{opt}),R(g_{mm})$.

If the modes are well matched, i.e. $g_{opt,i}=h_{opt,i}$, then there is only one accessible spatial mode. As long as the system has no non-collective excitations, we may also rewrite more familiar forms $\Sigma_{re}=R^\dagger E$, $N_e=E^\dagger E$, $N_r=R^\dagger R$. Then the Hamiltonian is simply
\begin{multline}
    H=\delta_{a}\hat{a}^{\dagger}\hat{a}+\delta_{b}\hat{b}^{\dagger}\hat{b}+\delta_{e} E^\dagger E +\delta_{r} R^\dagger R\\
    +g_{opt,c} \hat{a} E^\dagger + \Omega_b R^\dagger E + g_{mm,c} \hat{b} R^\dagger + h.c.
\end{multline}

Finally, defining $g_{opt,c}\equiv \sqrt{N} g_{opt}$, where $g_{opt}$ is the root mean square optical coupling across the cloud, this reduces to Eqn.~\ref{eqn:ham}.

\subsection{Alternate approach to derive a simplified Hamiltonian}

Going back to the time independent full Hamiltonian of the system

\begin{multline}
\label{eqn:full ham}
H=\delta_{a}\hat{a}^{\dagger}\hat{a}+\delta_{b}\hat{b}^{\dagger}\hat{b}+\sum_{i}\multiL\delta_{e}\hat{\sigma}_{ee}^{i}+\delta_{r}\hat{\sigma}_{rr}^{i}+\delta_{f}\hat{\sigma}_{ff}^{i}\\
+\biggl(g_{opt,i}\hat{a}\hat{\sigma}_{eg}^{i}+\Omega_{b,i}\hat{\sigma}_{re}^{i}+g_{mm}\hat{b}\hat{\sigma}_{fr}^{i}+\Omega_{uv,i}\hat{\sigma}_{fg}^{i}+h.c.\biggr)\biggr]
\end{multline}

As before, the subscripts $g$, $e$, $r$, and $f$ represent the $5\textrm{S}_{1/2}$, $5\textrm{P}_{3/2}$, $36\textrm{S}_{1/2}$, and $35\textrm{P}_{1/2}$ states respectively. Here we have assumed no position dependent phase and amplitude variation for the mmwave field across the atomic cloud, leading to a uniform coupling given by $g_{mm}$. This is reasonable since the cloud is much smaller than size of the mmwave mode. The amplitude and phase variations of for the $780$ mode $a$, and the blue and the UV couplings are contained in the respective couplings. For the moment we ignore the standing wave nature of the field $a$, and assume that $g_i = |g_i|e^{i \bf{k_{780}.r_i}}$, as for a running wave. We will reintroduce the effect of standing wave after simplifying the problem and clarifying the formalism. 

We now ansatz collective excitation operators which create collective excitations in the respective atomic states : 
\begin{align}
\label{eqn:collective states}
\begin{split}
E^{\dag} &= \frac{1}{\sqrt{\sum_{j}|g_j|^2}}\sum_{j} |g_j| e^{i \bf{k_{780}.r_j}} \hat{\sigma}_{eg}^{j} \\
R^{\dag} &= \frac{1}{\sqrt{\sum_{j}|g_j|^2}}\sum_{j} |g_j| e^{i \bf{(k_{780}+k_{481}).r_j}}  \hat{\sigma}_{rg}^{j}\\
F^{\dag} &= \frac{1}{\sqrt{\sum_{j}|g_j|^2}}\sum_{j}|g_j| e^{i \bf{(k_{780}+k_{481}).r_j}} \hat{\sigma}_{fg}^{j}
\end{split}
\end{align}
Note that these operators are all defined with respect to excitation from the manybody state with all the atoms in the ground, $5\textrm{S}_{1/2}$, state, which we treat as our vacuum. In defining the ${R}$ and ${F}$ operators, we have not included any intensity variation of the blue beam. This is in anticipation of the assumption that the waist of the blue beam is large compared to the $780$~nm mode waist, which is satisfied for our system. \emph{Without the UV drive}, the Hamiltonian in equation~\ref{eqn:full ham} can be rewritten as:

\begin{multline}
\label{eqn:collective ham-no uv}
H=\delta_{a}\hat{a}^{\dagger}\hat{a}+\delta_{b}\hat{b}^{\dagger}\hat{b}+\delta_{e} E^{\dag} E+\delta_{r} R^{\dag} R+\Delta F^{\dag} F \\
+\left(g_{opt}\sqrt{N}\hat{a} E^{\dag}+\Omega_{b} E R^{\dag}+g_{mm} \hat{b}^{\dag} F^{\dag}R+h.c.\right)
\end{multline}

where we have identified $\delta_f$ with $\Delta$ from the main text, $g_{opt} = \sqrt{\frac{\sum_{j}|g_j|^2}{N}}$ is the root mean square of the optical couplings across the atomic cloud, $N$ is the number of atoms, and $\Omega_b$ is the homogeneous blue Rabi frequency. Now we make the key assumption, that the operators $E$,$R$, and $F$ obey bosonic commutation relations. This can be seen from a Holstein-Primakoff type treatment~\cite{holstein-primakoff}, and is valid as long as the excitation numbers are small. Note that, in general, inhomogeneous couplings lead to excitations out of the bosonic manifold, but here we ignore that because we are in the low excitation regime, as well as the fact that those couplings are not collectively enhanced.

In the limit of high single atom cooperativity on the mmwave transition, the last term of the Hamiltonian in equation~\ref{eqn:collective ham-no uv} adds a strong non-linearity to the system. This is precisely the Jaynes-Cummings like non-linearity that makes this system attractive for generating strong all-to-all interactions between atoms.

To introduce the UV driving, we define a new collective operator
\begin{align*}
\overline{F}^{\dag} = \frac{1}{\sqrt{\sum_{j}|\Omega_{uv,j}|^2}}\sum_{j} |\Omega_{uv,j}| e^{i \bf{k_{297}.r_j}} \hat{\sigma}_{fg}^{j}
\end{align*}

Now the coherent UV drive term in the full Hamiltonian~\ref{eqn:full ham} is simply $(\sqrt{N}\Omega_{uv} \overline{F}^{\dag}+h.c)$, where $\Omega_{uv} = \sqrt{\frac{\sum_{j}|\Omega_{uv,j}|^2}{N}} $ is the root mean square of the UV couplings across the atomic cloud. It is now clear that if the $780$, blue and the UV fields are phase matched, $\bf{k_{780}+k_{481} = k_{297}}$, \emph{and} the $780$ field and the UV laser are mode-matched, $F$ and $\overline{F}$ are the \emph{same} operator, and the Hamiltonian~\ref{eqn:full ham} becomes

\begin{multline}
\label{eqn:collective ham}
H=\delta_{a}\hat{a}^{\dagger}\hat{a}+\delta_{b}\hat{b}^{\dagger}\hat{b}+\delta_{e} E^{\dag} E+\delta_{r} R^{\dag} R+\Delta F^{\dag} F \\
+\left(g_{opt}\sqrt{N}\hat{a} E^{\dag}+\Omega_{b} E R^{\dag}+g_{mm} \hat{b}^{\dag} F^{\dag}R+\sqrt{N}\Omega_{uv} F^{\dag}+h.c\right)
\end{multline}

Note the operator $F$ can only obey bosonic commutation relations to the extent that the UV does not create significant excitation in the $35\textrm{P}$ Rydberg state compared to the total atom number $N$. We have also ignored the depletion of the ground state. Both these assumptions are justified when $|\Omega_{uv}| \ll |\Delta|$, which is the regime we operate in. The detuning also adds an extra suppression factor to the leakage outside the bosonic manifold due to inhomogeneous couplings. 

This Hamiltonian can now be linearized in straight-forward manner by displacing the bosonic operator $F$ using the unitary transformation 
\begin{equation*}
U = \exp\left(\frac{\sqrt{N}\Omega_{uv}}{\Delta}F^{\dag}-\frac{\sqrt{N}\Omega^*_{uv}}{\Delta}F\right)
\end{equation*}
and neglecting the (small) non-linear $b^{\dag}F^{\dag}R$ term. The resulting Hamiltonian is:

\begin{multline}
\label{eqn:lin collective ham}
H_{lin}=\delta_{a}\hat{a}^{\dagger}\hat{a}+\delta_{b}\hat{b}^{\dagger}\hat{b}+\delta_{e} E^{\dag} E+\delta_{r} R^{\dag} R \\
+\left(g_{opt}\sqrt{N}\hat{a} E^{\dag}+\Omega_{b} E R^{\dag}+g_{mm}\frac{\sqrt{N}\Omega_{uv}}{\Delta} \hat{b}^{\dag} R+h.c.\right)
\end{multline}

This is equivalent to expanding the Hamiltonian around the steady state produced by the coherent UV drive and neglecting the small non-linearity. 
\subsection{Heisenberg-Langevin equations}
We start by writing the Heisenberg-Langevin equations~\cite{gardiner1985input} for non-linear Hamiltonian in equation~\ref{eqn:collective ham}, partly to obtain another perspective on linearization:
\begin{equation}
\label{eqn:langeqns}
\begin{split}
\dot{a}&=(i\delta_a-\frac{\kappa_{opt}}{2})a - i\sqrt{N} g_{opt} E+\sqrt{\kappa^{ext}_{o}}a_{in}(t)\\
\dot{E}&=(i\delta_e-\frac{\Gamma}{2})E - i\sqrt{N} g_{opt} a - i\Omega_b R\\
\dot{R}&=(i\delta_r-\frac{\Gamma_R}{2})R - i \Omega_b^* E+i g_{mm}b F\\
\dot{b}&=(i\delta_b-\frac{\kappa_{mm}}{2})b - i g_{mm}R F^{\dag}+\sqrt{\kappa^{ext}_{mm}}b_{in}(t)\\
\dot{F}&=(i\Delta-\frac{\Gamma_{F}}{2})F - i g_{mm}R b^{\dag}-i\sqrt{N}\Omega_{uv}
\end{split}
\end{equation}

where we have introduced the decay rates of the cavities ($\kappa_{opt}$ and $\kappa_{mm}$), the input fields for the optical and mmwave modes ($a_{in}$ and $b_{in}$), the incoupling rates for the optical and mmwave inputs  ($\kappa^{ext}_o$ and $\kappa^{ext}_{mm}$) and the dephasing/decay rates of the various collective states ($\Gamma$,$\Gamma_R$, and $\Gamma_F$). The last three equations contain products of operators which lead to the non-linearity in these equations. The non-linear term in the last equation can be neglected as long as the collective UV drive is stronger than the mmwave coupling, which is true as long as there is not a strong mmwave drive or a large population in the $36\textrm{S}_{1/2}$ state. The operator $F$, can now be adiabatically eliminated by setting $\langle\dot{F}\rangle = 0$, yielding $\langle F \rangle = i\frac{\sqrt{N}\Omega_{uv}}{\Delta-i\Gamma_F/2}$
It is clear that as long as $|\Delta|\gg\Gamma_F/2$, this dephasing is inconsequential, so we drop it. Replacing the operator $F$ by its average value leads to the required linearized Heisenberg-Langevin equations, which could also have been directly obtained from the linearized Hamiltonian~\ref{eqn:lin collective ham}.

At the dark polariton resonance, i.e. with $\delta_a=\delta_b=\delta_e=\delta_r=0$, equations~\ref{eqn:langeqns} may be rewritten in matrix form after linearization:
\def\psi{
\begin{bmatrix}
    a \\
    E \\
    R \\
    b
\end{bmatrix}}

\def\Hmat{
\begin{bmatrix}
    i\frac{\kappa_{opt}}{2} & G_o & 0 & 0 \\
    G_o & i\frac{\Gamma}{2} & \Omega_b & 0 \\
    0 & \Omega_b & i\frac{\Gamma_R}{2} & G_{mm}\\
    0 & 0 & G_{mm} & i\frac{\kappa_{mm}}{2}
\end{bmatrix}}

\def\u{
\begin{bmatrix}
    \kappa^{ext}_{o}a_{in}(t) \\
    0 \\
    0 \\
    \kappa^{ext}_{mm}b_{in}(t)
\end{bmatrix}}

\begin{equation}
\frac{d}{dt}\psi=i \Hmat \psi+\u
\end{equation}
Here we have defined $G_o= g_{opt}\sqrt{N}$,$G_{mm}=g_{mm}\sqrt{N}\frac{\Omega_{uv}}{\Delta}$. We can now transform to Fourier space, which reduces the problem to solving linear algebraic equations 
\begin{equation}
\label{eqn:HLfourier}
\bf{0=G(\omega)v(\omega)+u(\omega)}
\end{equation}

where we have defined 
\begin{align*}
\bf{v(\omega)} &= \begin{bmatrix} 
a(\omega) & E(\omega) & R(\omega) & b(\omega)
\end{bmatrix}^T \\
\bf{u(\omega)} &= \begin{bmatrix}
    \kappa^{ext}_{o}a_{in}(\omega) & 0 & 0 & \kappa^{ext}_{o}a_{in}(\omega)
\end{bmatrix}^T \\
\bf{G(\omega)}  &= \begin{bmatrix}
    (i\omega-\frac{\kappa_{opt}}{2}) & G_o & 0 & 0 \\
    G_o & (i\omega-\frac{\Gamma}{2}) & \Omega_b & 0 \\
    0 & \Omega_b & (i\omega-\frac{\Gamma_R}{2}) & G_{mm}\\
    0 & 0 & G_{mm} & (i\omega-\frac{\kappa_{mm}}{2})
\end{bmatrix}
\end{align*}
\subsection{Transducer transfer function and conversion efficiency}
The equations~\ref{eqn:HLfourier} to yield $a(\omega)$ in terms of $a_{in}(\omega)$ and $b_{in}(\omega)$. For mmmwave to optical conversion, we neglect the $a_{in}$ term since there is no appreciable thermal input on the optical side. Noting that the input-output relation is $a_{out}(\omega) = \sqrt{\kappa_{opt}^{ext}} a$, we can write in general

\begin{align}
\begin{split}
    a_{out}(\omega) = S(\omega)b_{in}(\omega)\\
    \langle a_{out}^{\dag}(\omega) a_{out}(\omega)\rangle = |S(\omega)|^2\langle b_{in}^{\dag}(\omega) b_{in}(\omega)\rangle
\end{split}
\end{align}

where $S(\omega)$ can be interpreted as a transfer function for the transducer between from one of the input ports of the mmwave cavity to one of the output ports of the optical cavity, and $|S(\omega)|^2$ defines a conversion efficiency. In general it can be shown that
\begin{equation}
    S(\omega) = \sqrt\frac{\kappa_{opt}^{ext}\kappa_{mm}^{ext}}{\kappa_{opt}(\omega)\kappa_{mm}(\omega)}\frac{2\sqrt{C_o(\omega)C_{mm}(\omega)C_b(\omega)}e^{i\phi}}{(C_b(\omega)+(1+C_o(\omega))(1+C_{mm}(\omega)))}
\end{equation}

where we have defined $\gamma_j(\omega) = \gamma_j-2 i \omega$ for $\gamma_j \in \{\kappa_{opt},\kappa_{mm},\Gamma,\Gamma_{R}\} $, $\phi$ is the phase difference between the blue and the UV beams in the rotating frame, and $C_o(\omega) = \frac{4 |G|^2}{\kappa_{opt}(\omega)\Gamma(\omega)}$, $C_b(\omega) = \frac{4|\Omega_b|^2}{\Gamma(\omega)\Gamma_R(\omega)}$, and $C_{mm}(\omega) = \frac{4 |G_{mm}|^2}{\kappa_{mm}(\omega)\Gamma_R(\omega)}$ are generalized frequency dependent cooperativities. On resonance, the conversion efficiency is simply

\begin{equation}
|S(0)|^2 = \frac{\kappa_{o}^{ext}\kappa_{mm}^{ext}}{\kappa_{o}\kappa_{mm}} \frac{4 C_{o}C_b C_{mm}}{(C_b+(1+C_{mm})(1+C_o))^2}
\end{equation}
where the generalized cooperativities are evaluated at $\omega=0$ and simplify to those defined in the main text, resulting in full expression reducing to the internal conversion efficiency for single ended cavities.

Since we measure count rates at the output of the optical cavity, we want to calculate $\langle a_{out}^{\dag}(t)a_{out}(t)\rangle$, which is the mean photon flux at the output at time t. By definition $a_{out}(t)=\frac{1}{\sqrt{2\pi}}\int_{-\infty}^{\infty}a_{out}(\omega)e^{-i\omega t}d\omega$, allowing us to calculate \emph{non-normalized} first order correlation functions:

\begin{multline}
    \overline{g}^{(1)}(\tau) = \langle a_{out}^{\dag}(t+\tau)a_{out}(\tau) \rangle\\
    = \frac{1}{2\pi}\iint \limits_{-\infty}^{+\infty}S(\omega)^*S(\omega') \langle b_{in}^{\dag}(\omega)b_{in}(\omega')\rangle e^{-i\omega \tau} d\omega d\omega'
\end{multline}

For our experiments the input field is a coherently displaced thermal field, which we write as
\begin{equation*}
b_{in}(t) = \beta e^{-i\omega_D t} + b_{in}^{th}(t) 
\end{equation*}
 and consequently
\begin{equation*}
b_{in}(\omega) = 2\pi \beta \delta(\omega-\omega_D) + b_{in}^{th}(\omega)
\end{equation*}
, where we have separated the coherent and the thermal parts and $|\beta|^2$ is the mean photon flux of the coherent drive and in this rotating frame, $\omega_D$ is the detuning of the drive from the mmwave cavity. For a thermal input, $\langle {b_{in}^{th}}^{\dag}(\omega)b_{in}^{th}(\omega')\rangle = n_{th}\delta(\omega-\omega')$~\cite{loudon2000quantum}, $n_{th}$ is the number of thermal photons at frequency $\omega$, which we have assumed to be constant for the narrow transduction bandwidth we care about.

A slightly subtle point is that a thermal drive is coupled through all the input ports of the cavity, which includes the inner walls, while a coherent drive is coupled only through the external port. In effect this leads to the following result:

\begin{multline}
    \overline{g}^{(1)}(\tau) = \overline{g}_{coh}^{(1)}(\tau)+\overline{g}_{th}^{(1)}(\tau)\\
    = |\beta|^2 |S(\omega_D)|^2 e^{-i \omega_{D} \tau}
    +\frac{\kappa_{mm}}{\kappa_{mm}^{ext}} \frac{n_{th}}{2\pi} \int_{-\infty}^{\infty}|S(\omega)|^2 e^{-i \omega \tau} d\omega
\end{multline}

where the $\frac{\kappa_{mm}}{\kappa_{mm}^{ext}}$ factor in the second term accounts for the fact that we have defined our transfer function from the external port in which the coherent drive is coupled. One can use similar methods to compute the second order correlation function $g^{(2)}(\tau)$ and prove equation~\ref{eqn:full g2} in Methods.

One remaining thing is to draw a connection between the number of coherently driven intra-cavity photons, which we measure from the weak dispersive shift of the polariton, and $\beta$. It is trivial to solve the Langevin equations for just the mwmave cavity with only a coherent drive, leading to
\begin{equation}
n_{ph}= \langle b^\dag(t) b(t) \rangle = \frac{4|\beta|^2 \kappa_{mm}^{ext}}{{\kappa_{mm}}^2}
\end{equation}

The rate of photons, $R$, at the output of the optical cavity is given by $\overline{g}^{(1)}(0)$, which can now be written as:
\begin{multline}
    R = R_{coh} + R_{th}\\ =  \frac{n_{ph}\kappa_{mm}^2}{4 \kappa_{mm}^{ext}}|S(\omega_D)|^2
    +\frac{\kappa_{mm}}{\kappa_{mm}^{ext}} \frac{n_{th}}{2\pi} \int_{-\infty}^{\infty}|S(\omega)|^2 d\omega
\end{multline}
Note that since $|S(\omega)|^2$ is proportional to $\kappa_{mm}^{ext}$, the output rate is independent of the external coupling rate when parametrized in terms of the intra-cavity photon number. Measuring the rate $R$, with a calibrated $n_{ph}$ coherent drive and a rate $R_{th}$ without the coherent drive allows us to deduce the internal conversion efficiency:
\begin{equation}
\eta_{o\leftrightarrow mm} = \frac{1}{f_o}\frac{\kappa_{opt} }{\kappa_{opt}^{ext}}\frac{4 (R-R_{th})}{\kappa_{mm} n_{ph}}
\end{equation}
where $f_o$ is the efficiency of the optical path after the optical cavity.

\subsection{Effect of a standing wave cavity}
\label{SI:standing wave}
So far we have neglected the fact that the optical cavity mode $a$ is that of a standing wave, and we have only considered collective states with one of the running wave phase ``imprints" in definitions~\ref{eqn:collective states}. The standing wave also couples to other collective states with the other running wave phase:

\begin{align*}
\begin{split}
E'^{\dag} &= \frac{1}{\sqrt{\sum_{j}|g_j|^2}}\sum_{j} |g_j| e^{i \bf{-k_{780}.r_j}} \hat{\sigma}_{j}^{eg} \\
R'^{\dag} &= \frac{1}{\sqrt{\sum_{j}|g_j|^2}}\sum_{j} |g_j| e^{i \bf{(-k_{780}+k_{481}).r_j}}  \hat{\sigma}_{j}^{rg}\\
F'^{\dag} &= \frac{1}{\sqrt{\sum_{j}|g_j|^2}}\sum_{j}|g_j| e^{i \bf{(-k_{780}+k_{481}).r_j}} \hat{\sigma}_{j}^{fg}
\end{split}
\end{align*}

Note that the UV beam \emph{only} couples to the state created by the $F^\dag$ operator and and not the newly defined $F'^\dag$ operator, because of orthogonal phase profiles. We also neglect the mmwave cavity coupling of this ``$F'$ state" to the one created by the $R'^\dag$ operator since it is both off-resonant and not collectively enhanced. Effectively, this leads to two more terms in the Hamiltonian~\ref{eqn:lin collective ham} giving:
\begin{equation*}
H_{lin,stand} = H_{lin}+\left(g_{opt}\sqrt{N}\hat{a} E'^{\dag}+\Omega_{b} E' R'^{\dag}+h.c.\right)
\end{equation*}

This just adds two more Langevin equations for the operators $E'$ and $R'$, which can still be solved in closed form. Essentially it amounts to the same set of Langevin equations in the fourier space as before, but now with the replacement 
\begin{equation*}
\kappa_{opt}(\omega)  \rightarrow \kappa_{opt}(\omega)\left(1+\frac{C_o(\omega)}{1+C_b(\omega)}\right) 
\end{equation*}
Physically this describes loss of an optical photon by atomic absorption of the ``wrong" running wave, but this loss is still protect by the EIT, so in the limit of high blue power it becomes inconsequential.

A final key point is that we have so far defined $g_{opt}$ as the coupling between one of the running waves and atoms. In a standing wave cavity, in terms of this coupling, the vacuum Rabi splitting that this results in is $2 \sqrt{N} \sqrt{2}g_{opt}$ instead of $2 \sqrt{N} g_{opt}$. This is important for correctly calculating the effective atom number.

\subsection{Interconversion as happening between the dark polariton and mmwaves}
\label{SI:interconversion dark polariton }
We mentioned in the main text that it in the large optical cooperativity limit, it is possible to interpret the interconversion as happening between the mmwave mode and the dark polariton. Here we show that explicitly. We start by applying a unitary transformation that maps the operators in the Hamiltonian~\ref{eqn:lin collective ham} as follows :

\begin{multline}
\label{eqn:operator transform}
\begin{split}
a &= \frac{1}{\sqrt{2}}(B_++B_-) \sin{\theta_D} + D \cos{\theta_D}\\
E &= \frac{1}{\sqrt{2}}(B_+-B_-)\\
R &=  \frac{1}{\sqrt{2}}(B_++B_-) \cos{\theta_D} - D \sin{\theta_D}\\
\end{split}
\end{multline}
where $D$, $B_+$, and $B_-$ are dark polariton, upper bright polariton and lower bright polariton destruction operators respectively, and $\tan{\theta_D}=g/\Omega$ is the dark polariton rotation angle. At the resonant point, making the substitutions~\ref{eqn:operator transform}, and projecting the Hamiltonian~\ref{eqn:lin collective ham} into the dark polariton manifold results in 
\begin{equation*}
    P_DH_{lin}P_D = \left( - g_{mm}\sin{\theta_D}\frac{\Omega_{uv}}{\Delta} b^\dagger D   + h.c.\right) 
\end{equation*}
which is simply a beam-splitter between the mmwave mode and the dark polariton. While for our system, we cannot entirely ignore the effect of off-resonant bright polariton excitation to get quantitative agreement, this analysis provides a simple approximate interpretation of the process.

\section{Dispersive shift of the dark polariton due to a mmwave drive}
\label{SI:dispersive shift}
To measure the internal conversion efficiency, we calibrate our mmwave drive strength by measuring the energy shift of the dark polariton in the weakly dispersive regime, with an atom-mmwave cavity detuning of $\Delta = 2\pi\times1.4$~MHz, much larger than $g_{mm}$. Even though the non-linearity is weak, we probe the dark polariton with a very weak optical drive to ensure that on average there is much less than a single atomic excitation in the system. In the dispersive limit, the effect of the mmwave drive is to ac Stark shift the $36\textrm{S}$ atomic state by $(n_{mm}+1) g^2/\Delta$, where $n_{mm}$ would the average number of photons in the bare mmwave cavity for a given drive strength and the added thermal occupation. Since the thermal occupation (as well as the vacuum coupling) is constant, any shift we measure on the top of that is simply proportional to the number of drive photons, $n_{ph}$. At the lowest order in perturbation theory, the dark polariton experiences a shift that is set by the fraction of Rydberg state admixture in its composition: $\delta(n_{ph}) = n_{ph} g^2 \sin^2{\theta_D}/\Delta$. 

To build confidence in this calibration, we model the shift by solving the master equation for the non-linear Hamiltonian, Eqn.~\ref{eqn:collective ham-no uv} including $0.6$ thermal photon population expected at $5$~K with an added coherent drive that by itself would drive an average of $n_{ph}$ photons into the cavity at steady state. Our master equation simulations show that this simple expression slightly overestimates the shift due to higher order corrections, primarily because of the slight admixing of the $35\textrm{P}$ state. While in principle we can just use the master equation simulation to predict $n_{ph}$, it is possible to obtain a much more accurate analytical expression for the dark polariton shift by calculating the eigenvalues of the Hamiltonian~\ref{eqn:collective ham-no uv} in the single optical/atomic excitation and $n_{mm}$ mmwave excitation manifold. In that limited manifold and on resonance for the optical cavity and the blue transition, the hamiltonian can be written as:

\begin{equation*}
H(n_{mm}) = \begin{bmatrix}
    0 & G_o & 0 & 0 \\
    G_o & 0 & \Omega_b & 0 \\
    0 & \Omega_b & 0 & \sqrt{n_{mm}+1}g_{mm}\\
    0 & 0 & \sqrt{n_{mm}+1}g_{mm} & \Delta
\end{bmatrix}
\end{equation*}

This Hamiltonian would lead to a fourth-order eigenvalue equation. Since we are only interested in the eigenvalue near 0 (which is where the dark polariton starts out) and small shifts, the higher order terms in the eigenvalue equation can be neglected. Keeping only the first order term leads to the previously mentioned simple expression. Going up to the second order yields the following expression for the dark polariton shift:

\begin{equation}
    \delta(n_{mm}) = f\left(\frac{\Delta-sgn(\Delta) \sqrt{\Delta^2+\frac{4(n_{mm}+1)g_{mm}^2\sin^2{\theta_D}}{f}}}{2}\right)
\end{equation}

where $f$ is a correction factor given by 

\begin{equation*}
    f = \frac{G_o^2+\Omega_b^2}{G_o^2+\Omega_b^2+(n_{mm}+1)g_{mm}^2}
\end{equation*}

In general the shift that we observe is given by \linebreak $\delta(n_{ph}+n_{th})-\delta(n_{th})$, where $n_{th}$ is the number of thermal photons which, for calulating $n_{ph}$ from this shift, we set to $0.6$.

\section{Quasi-phase matching}
\label{SI:qpm}
We have observed that the high powered focused UV laser can significantly affect the coatings of the optical cavity and in instances of steady state turn on, permanently damage them in a few seconds. In our current setup, phase matching requires that UV beam co-propagate with the optical cavity mode, i.e. in addition to energy conservation in a four-wave mixing process, there is also a momentum conservation requirement. This limits us to small duty cycles with the UV beam, lest we damage the optical cavity. In future this issue can be entirely avoided through quasi-phase matching, which would allow the blue and UV beams to be sent from any direction as long as they are co-propagating with each other.

In vacuum, the linear dispersion of light means that, so long as all beams are co-/counter-propagating as necessary, momentum conservation (also called ``phase matching'' in this context) occurs automatically. If either (a) the light is propagating in a dispersive medium or (b) geometric or other constraints make it favorable to choose different beam directions, the ``missing momentum'' has to come from somewhere. In the language of collective enhancements, any residual momentum is left in the recoil of the electron/atom that went around the four-wave-mixing loop, resulting in which-path information that destroys the collective enhancement of the transduction process.

The technique of quasi-phase-matching takes the missing momentum from periodic structures written into the four-wave mixing medium. In the solid state, this may be achieved by periodic poling of the nonlinear crystal, which leaves the index of refraction unchanged but reverses the sign of, for example, the $\chi^{(2)}$ nonlinear coefficient. Matching this poling period to the k-vector mismatch in the mixing process provides the necessary momentum. This may alternatively be understood as Umklapp scattering off of the lattice imposed by the periodic poling.

In the context of our transducer, the density of atoms is sufficiently low that the effective index of refraction at all wavelengths of interest is essentially unity, so the only application of quasi-phase matching is to enable selection of more convenient directions for the various laser fields. In particular, it can be convenient to avoid sending the UV laser through the optical mirrors, which it can damage. To achieve such periodic poling, the atoms could be localized in optical lattices whose wavelengths provide the necessary quasi momentum. In particular, we propose to fully relax all phase matching requirements by:

\begin{itemize}
    \item Trapping the atoms in an intracavity optical lattice at $1560$~nm, which therefore has a spacing of $1560/2=780$~nm, and thus can exactly compensate from all recoils associated with collectively absorbing/emitting photons from/into the optical resonator
    \item Sending $297$~nm and $481$~nm beams in exactly co-propagating with one another (with any orientation to the $780$~nm cavity-- whatever is most convenient), and trapping the atoms in a co-aligned lattice with a wavelength of $1560$~nm to again absorb all of the quasi-momentum.
\end{itemize}

This fully-quasi-phase matched setup essentially allows the $481$~nm and $297$~nm beams can come in from any direction, and potentially even be enhanced in their own separate buildup cavities.

\section{Near-Resonant Four-Wave Mixing}
\label{SI:fwm}
In traditional nonlinear optics, $\chi^{(2)}$, $\chi^{(3)}$ nonlinear processes in (for example) Lithium Niobate are performed at extremely large detunings from the electronic resonances~\cite{boyd2020nonlinear}, at the expense of extremely high laser power requirements. Here we are able to operate at moderate laser powers on the very weak $5\textrm{S}_{1/2}\leftrightarrow 35\textrm{P}_{1/2}$ and $5\textrm{P}_{3/2}\leftrightarrow 36\textrm{S}_{1/2}$ transitions, achieving high conversion efficiencies and large transduction bandwidths. This is because we are able to work on or \emph{very near} all of the atomic resonances.

In solid state nonlinear optics it is dangerous to operate near the electronic lines because they are strongly inhomogeneously broadened, making loss-suppression mechanisms like EIT, which are typically detuning-independent in homogeneously broadened systems, turn off near the atomic lines. By contrast, our Rydberg lines are extremely narrow compared to other physics in the system, enabling us to work directly on resonance.

The other concern with near-resonant operation is the presence of substantial Rydberg population, which could lead to both blockade effects~\cite{urban2009observation,gaetan2009observation} and plasma formation~\cite{robinson2000spontaneous}. Because our atom cloud is quite dilute, and because we are working at relatively low principal quantum numbers, this is also not a concern in our apparatus. Note also that the $36\textrm{S}_{1/2}$ population is at most the number of photons currently being interconverted, and so this is not likely to pose a limitation. The $35\textrm{P}_{1/2}$ population is more likely to be problematic, as we employ UV-dressing to this state in order to achieve collective enhancement of a factor of 2 on the mmwave cooperativity (2 atoms on average dressed to the $35\textrm{P}$ state); here the low density of the atomic cloud and small $C_6$ coefficient (low principal quantum number) protect us.

\setcounter{secnumdepth}{2}
	
\end{document}